\documentclass[aip, jcp, notitlepage, floatfix, reprint, citeautoscript, twocolumn]{revtex4-1}

\usepackage{amsmath}
\usepackage{amsfonts}
\usepackage{graphicx}
\usepackage{dcolumn}
\usepackage[table]{xcolor}
\usepackage[version=3]{mhchem}
\usepackage{transparent}
\usepackage{gensymb}
\usepackage{tabularx}
\usepackage{etoolbox}
\usepackage[hidelinks]{hyperref}
\usepackage{bm}
\usepackage{relsize}
\usepackage[normalem]{ulem}

\def\maxfloatwidth{%
  \ifdim\columnwidth>246.0pt
  300.0pt  \else
  \columnwidth
  \fi
}

\newcommand{\tbf}[1]{\textbf{#1}}

\newcommand{\mrm}[1]{\mathrm{#1}}
\newcommand{\mbf}[1]{\mathbf{#1}}
\newcommand{\tcr}[1]{\textcolor{black}{#1}}
\newcommand{\tcb}[1]{\textcolor{black}{#1}}
\newcommand{\etal}{\emph{et al.}}

\begin{document}

\title{Dielectric response of thin water films: A thermodynamic perspective}

\author{Stephen J. Cox}
\affiliation{Yusuf Hamied Department of Chemistry, University of
  Cambridge, Lensfield Road, Cambridge CB2 1EW, United Kingdom.}
\email{sjc236@cam.ac.uk}

\author{Phillip L. Geissler}
\affiliation{Chemical Sciences Division, Lawrence Berkeley National
  Laboratory, Berkeley, CA 94720, United States.}
\affiliation{Department of Chemistry, University of California,
  Berkeley, CA 94720, United States.}
\email{geissler@berkeley.edu}

\date{\today}

\begin{abstract}
  The surface of a polar liquid presents a special environment for the
  solvation and organization of charged solutes, which differ from
  bulk behaviors in important ways. These differences have motivated
  many attempts to understand electrostatic response at aqueous
  interfaces in terms of a spatially varying dielectric permittivity,
  typically concluding that the dielectric constant of interfacial
  water is significantly lower than in the bulk liquid.  Such
  analyses, however, are complicated by the potentially nonlocal
  nature of dielectric response over the short length scales of
  interfacial heterogeneity.  Here we circumvent this problem for thin
  water films by adopting a thermodynamic approach.  Using molecular
  simulations, we calculate the solvent's contribution to the
  reversible work of charging a parallel plate capacitor. We find good
  agreement with a simple dielectric continuum model that assumes bulk
  dielectric permittivity all the way up to the liquid's boundary,
  even for very thin ($\sim 1$ nm) films. This comparison requires
  careful attention to the placement of dielectric boundaries between
  liquid and vapor, which also resolves apparent discrepancies with
  dielectric imaging experiments.
\end{abstract}

\maketitle

\section{Introduction}
\label{sec:intro}

Interest in confined water has exploded over the last decade or so,
owing principally to advances in the fabrication of devices at the
nanoscale \cite{geim2013van,celebi2014ultimate,jain2015heterogeneous},
the potential implications for `blue energy' and desalination
\cite{bocquet2020nanofluidics}, and as means to understand fundamental
properties of water \cite{algara2015square,fumagalli2018anomalously}
and its solutions \cite{nakamuro2021capturing,wang2021microscopic}. An
obvious consequence of the decreasing length scales associated with
confinement is an increase in the surface-to-volume ratio of liquid
water, which typically amplifies surface-specific effects relative to
large sample geometries. The notion of nanoconfined liquid water thus
having properties that are inherently different to its bulk
counterpart has inspired many attempts to reformulate intensive
material parameters typically used to describe the bulk fluid. In
particular, many years of investigation along these lines
\cite{ballenegger2005dielectric,bonthuis2011dielectric,bonthuis2012profile,schlaich2016water,loche2018breakdown,zhang2018note,olivieri2021confined,matyushov2021dielectric,mondal2021anomalous}
has focused on the static dielectric constant $\epsilon_{\rm liq}$,
whose role in mediating electrostatic interactions impacts upon, e.g.,
solvation, capacitance and electrokinetics. Further motivation for
such theoretical studies comes from recent dielectric imaging
experiments \cite{fumagalli2018anomalously} of water confined between
two atomically flat walls separated by distances as small as
0.8\,nm. These imaging results were inferred to report an interfacial
dielectric constant $\epsilon_{\rm int}=2.1$ (relevant to an
interfacial region of thickness $\ell_{\rm int} \approx 7.5\pm
1.5$\,\AA) that dominates the capacitance of a thin water film.  This
value, typical of a bulk nonpolar liquid, signifies a dramatic
departure from the polarizability of bulk water, for which
$\epsilon_{\rm liq}\approx 80$.

At the microscopic level, it is well recognized that water's
interfaces exhibit local average properties that differ from the bulk
liquid, varying continuously with depth within a molecular length
scale $\ell_{\rm int}$ of the surface
\cite{WidomRowlinson}. Accordingly, many studies have aimed to
rationalize confined water's electrostatic response in terms of a
local dielectric constant $\epsilon(z)$ that varies with position $z$
along the surface normal
\cite{ballenegger2005dielectric,bonthuis2011dielectric,bonthuis2012profile,schlaich2016water,loche2018breakdown,loche2020universal,jalali2020out,hamid2021abnormal}.
Molecular simulations have estimated $\epsilon(z)$ either from
polarization fluctuations, or from response to external electric
fields; in either case this approach relies upon interpreting features
that have been resolved at a fine scale within a theoretical framework
appropriate for macroscopic dielectric materials. In this study, we
pursue a different approach. Specifically, we assess the ability of a
simple dielectric continuum theory (DCT)---whose dielectric
permittivity does not vary with depth $z$---to predict free energy
differences when water films are subjected to external fields. An
advantage of this approach is that it is rooted in thermodynamics,
which obviates the need to resolve fluctuations/response at the
microscopic level. We will show that simple DCT with
$\epsilon(z)=\epsilon_{\rm bulk}=\mathrm{const.}$ not only gives a
good description of water's dielectric response under confinement, but
it also outperforms models that suppose a lower dielectric constant at
the interface. Moreover, we also find that for films comprising just
one or two layers of water molecules this simple DCT remains a
remarkably reasonable approximation.  We show that our analyses are
broadly in line with the experimental observations reported in
Ref.~\onlinecite{fumagalli2018anomalously}.

The rest of the article is arranged as follows. In
Sec.~\ref{sec:OverviewDielectrics} we briefly review linear response
theory for dielectric fluids, calling into question the notion of a
permittivity that varies with position over microscopic scales.  In
Sec.~\ref{sec:dct-outline} we analyze the polarization of a confined
dielectric continuum under periodic boundary conditions, and derive a
finite size correction for the thermodynamics of charging up a
parallel plate capacitor.  In Sec.~\ref{eqn:ApplyToSim} we use
molecular simulations of simple point charge models to assess the
accuracy of this correction, and compare extrapolated results with DCT
predictions. We subsequently assess the performance of more
complicated models in Sec.~\ref{sec:OtherModels}. In
Sec.~\ref{sec:SmallW} we investigate the length scales at which DCT
begins to fail. The sensitivity of the effective dielectric constant
to the definition of film thickness is discussed in
Sec.~\ref{sec:ExpComp}. We summarize our findings in
Sec.~\ref{sec:Summary}.

\section{Brief overview of dielectrics}
\label{sec:OverviewDielectrics}

In macroscopic DCT, the polarization $\mbf{P}$ in a medium is related
to the total electric field $\mbf{E}$ by the constitutive relation
\cite{fulton1978long,fulton1978dipole,fulton1983theory,caillol1992asymptotic}
\begin{equation}
  \label{eqn:constit-nl}
  4\pi\mbf{P}(\mbf{r}) =
  \int\!\mrm{d}\mbf{r}^\prime\,\left[{\bm \epsilon}(\mbf{r},\mbf{r}^\prime)-\mbf{1}(\mbf{r},\mbf{r}^\prime)\right]\cdot\mbf{E}(\mbf{r}^\prime),
\end{equation}
where ${\bm \epsilon}$ is the dielectric tensor,
$\mbf{1}(\mbf{r},\mbf{r}^\prime)=\mbf{U}\delta(\mbf{r},\mbf{r}^\prime)$
with $\mbf{U}$ the unit tensor, $\delta(\mbf{r},\mbf{r}^\prime)$ is
Dirac's delta function, and the domain of integration is the volume
occupied by the medium. Equation~\ref{eqn:constit-nl} is a nonlocal
relationship between $\mbf{P}$ and $\mbf{E}$. There are two routes to
arrive at the more familiar local relationship for a homogeneous,
isotropic dielectric
\begin{equation}
  \label{eqn:constit}
  4\pi\mbf{P}(\mbf{r}) = (\epsilon-1)\mbf{E}(\mbf{r}).
\end{equation}
The first is to simply assert locality i.e., ${\bm
  \epsilon}(\mbf{r},\mbf{r}^\prime) =
\epsilon\mbf{1}(\mbf{r},\mbf{r}^\prime)$. The second, more formal
approach acknowledges the underlying molecular granularity, and
supposes that ${\bm \epsilon}(\mbf{r},\mbf{r}^\prime)$ is a
short-ranged function such that,
\(
{\bm \epsilon}(\mbf{r},\mbf{r}^\prime) \simeq 0
\)
for $|\mbf{r}-\mbf{r}^\prime|>\ell_\epsilon$. The characteristic
length $\ell_\epsilon$ is determined by molecular correlations, and
previous simulation studies suggest $\ell_\epsilon\approx 6$\,\AA
\cite{zhang2016computing,cox2020dielectric}. If $\mbf{E}$ varies
slowly over distances comparable to $\ell_\epsilon$, then the nonlocal
relation (Eq.~\ref{eqn:constit-nl}) reduces to the local one
(Eq.~\ref{eqn:constit}), with
\[
\tcb{\epsilon\mbf{U} = \int\!\mrm{d}\mbf{r}^\prime\,{\bm \epsilon}(\mbf{r},\mbf{r}^\prime).}
\]

Interfaces between different media are treated as infinitely sharp
boundaries within DCT. Any polarization in a medium then results in an
induced surface charge $\sigma_{\rm ind}(\mbf{x}) =
\mbf{P}(\mbf{x})\cdot\hat{\mbf{n}}(\mbf{x})$ that occupies a region of
infinitesimal thickness, where $\hat{\mbf{n}}(\mbf{x})$ is the local
surface normal. Such a scenario is, of course, an idealization of
physical reality \cite{JacksonJohnDavid1999Ce/J}; as discussed above,
liquid interfaces have a finite length scale $\ell_{\rm int}$, which
is on the order of 1\,nm---or a few molecular diameters---for liquid
water close to its triple point. The induced surface charge density is
then understood to result from a physical charge distribution which is
localized in the interfacial region, but with a thickness comparable
to $\ell_{\rm int}$. Molecular simulations suggest that such
interfacial charge distributions may vary rapidly along $z$
\cite{ballenegger2005dielectric,bonthuis2011dielectric,bonthuis2012profile,schlaich2016water,loche2018breakdown}. Local
dielectric constants obtained from simulation exhibit similar
structure.

\begin{figure}[tb]
  \includegraphics[width=8cm]{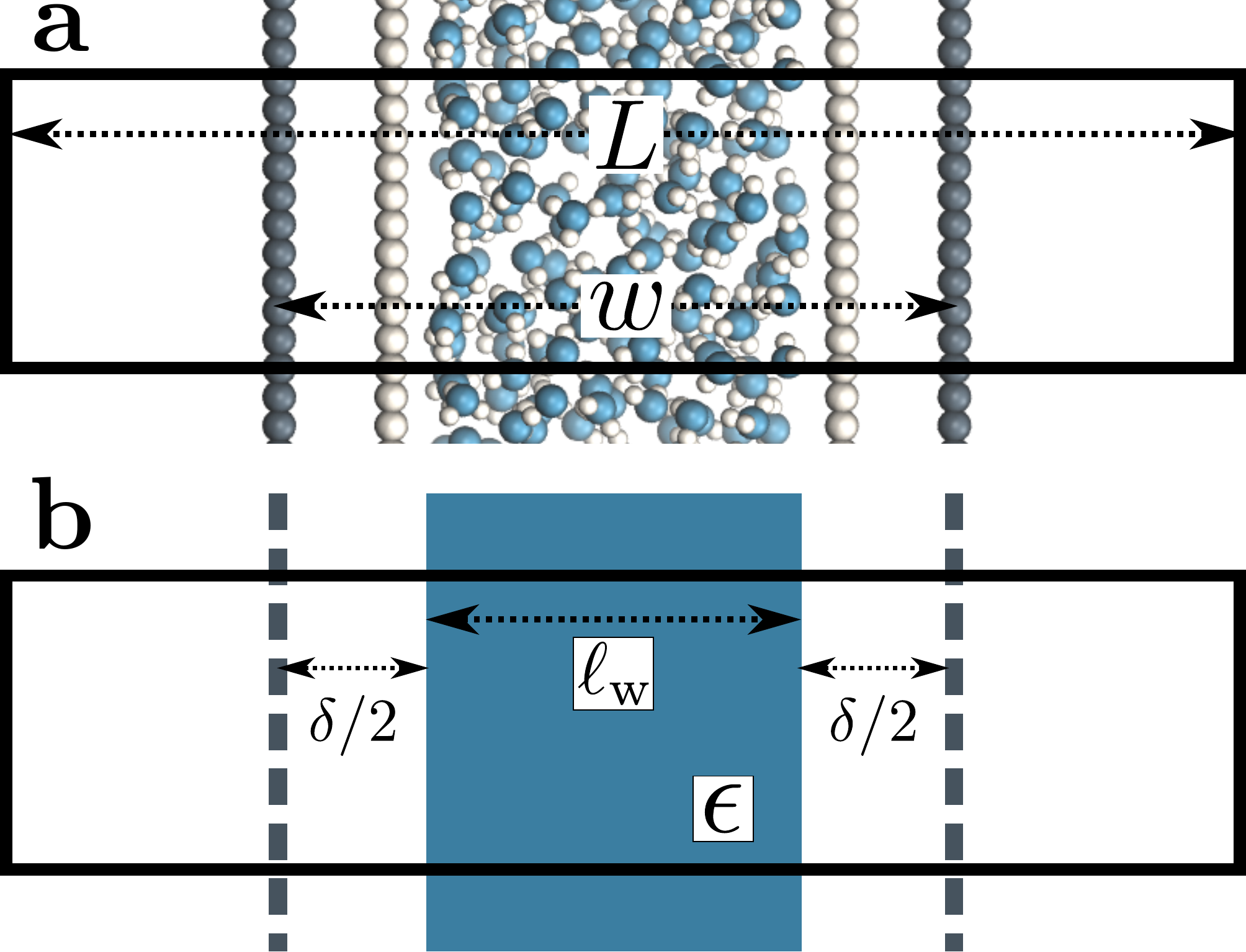}
  \caption{Molecular and continuum representations of the system
    considered. (a) Water molecules (oxygen atoms in blue) are
    confined between volume-excluding WCA particles (light gray). Dark
    gray circles represent point charges: negative on the left,
    positive on the right, separated by a distance $w$. (b) In the
    continuum representation, these planes of point charges are
    approximated as uniformly charged sheets, as indicated by the
    dashed dark gray lines. The effect of the WCA particles enters
    implicitly by bounding the solvent, itself represented as a
    continuum with dielectric constant $\epsilon$, within a slab of
    thickness $\ell_{\rm w}=w-\delta$, where $\delta/2$ indicates the
    distance between the solvent-vapor dielectric boundary, and the
    charged planes. In both (a) and (b), the simulation cell is
    periodically replicated in all three dimensions, and its length in
    the direction normal to the charged planes is $L$.}
  \label{fig:schematic}
\end{figure}

While it is reasonable to suppose that the properties of a material
may differ in regions close to the interface compared to those in
bulk, the notion of a local dielectric constant with variations on the
molecular scale is unsettling in a couple of respects. First, in going
from the nonlocal constitutive relation specified by
Eq.~\ref{eqn:constit-nl} to the local relation specified by
Eq.~\ref{eqn:constit}, we assumed that fields vary slowly over length
scales comparable to $\ell_\epsilon \approx \ell_{\rm int}$, so one
might therefore question the appropriateness of a local dielectric
constant. Second, even if one is content with the locality of
$\epsilon(z)$, DCT is a \emph{macroscopic} theory, and the
constitutive relations Eqs.~\ref{eqn:constit-nl} and~\ref{eqn:constit}
concern the $\emph{macroscopic}$ fields $\mbf{E}$ and
$\mbf{P}$. Obtaining these fields from the underlying microscopic
degrees of freedom thus requires a coarse graining procedure, and it
is reasonable to suppose that $\ell_\epsilon$ sets the minimum length
scale over which any such coarse graining should be performed.
\tcb{Local molecular response functions that vary rapidly in space are
  likely important for the solvation and spatial distribution of ions,
  as well as electrokinetic phenomena
  \cite{bonthuis2011dielectric,bonthuis2012profile}; it nonetheless
  remains} challenging to reconcile variations of $\epsilon(z)$ on the
molecular scale with this viewpoint of relating coarse grained
macroscopic fields (Eqs.~\ref{eqn:constit-nl}
and~\ref{eqn:constit}). By pursuing a thermodynamic perspective in
this paper, which directly compares predictions of simple DCT to free
energies obtained from molecular simulations, we avoid needing to
compute the macroscopic fields $\mbf{E}$ and $\mbf{P}$ from
microscopic degrees of freedom.

\section{Using simple DCT as a finite size correction}
\label{sec:dct-outline}

The extent of the physical systems we have in mind are microscopic in
one direction (perpendicular to the interface) but otherwise
macroscopic. To represent them in computer simulations, we take the
standard approach of imposing periodic boundary conditions in all
three Cartesian directions. Our simulated system is thus an infinite
stack of thin water slabs, separated by substantial but still
microscopic layers of vacuum, with an artificial periodicity. Because
electrostatic interactions are long in range, we anticipate
nonnegligible quantitative consequences of this periodicity,
particularly when its repeat length is not significantly larger than
the slab width.

To correct for such finite size effects, we adopt a strategy
previously used to assess system size-dependence for ion solvation in
similarly periodic slabs. Specifically, we extend our work in
Ref.~\onlinecite{cox2018interfacial} to develop a finite size
correction for the solvent contribution to the reversible work
required to charge up a parallel plate capacitor under periodic
boundary conditions (which we refer to as the `solvation' free energy,
$f_{\rm solv}^{(L)}$), based on the assumption that long-wavelength
solvent response underlying finite size effects is well-described by
DCT \cite{HunenbergerMcCammon1999sjc}. These predictions of DCT for
charging parallel plates that bound thin water slabs serve
simultaneously as a means to extrapolate computed free energies to the
thermodynamic limit, and also as a test of the assumptions underlying
DCT.

A representative snapshot of the system under consideration is shown
in Fig.~\ref{fig:schematic}a. The parallel plate capacitor is
approximated by two planes of $N_{\rm site}$ point charges arranged on
a square lattice, located at $z=\pm w/2$. The total charge of the
plane at $z=w/2$ is $Q=N_{\rm site}q_{\rm site}$, which is
equal-and-opposite to the plane at $z=-w/2$. The solvent water
molecules are confined between these two charged planes by tightly
packed volume-excluding Weeks-Chandler-Anderson \cite{WCA} (WCA)
particles (see Sec.~\ref{sec:Methods}). In most of what follows, the
WCA centers and the point charges coincide, though we will also
consider more general cases like those depicted in
Fig.~\ref{fig:schematic}a. We now make two continuum
approximations. First, water is treated as a dielectric slab with
dielectric constant $\epsilon$, spanning $z=-\ell_{\rm w}/2$ to
$z=+\ell_{\rm w}/2$, as indicated in Fig.~\ref{fig:schematic}b.  A
value of $\ell_{\rm w}$ appropriate to our molecular system is not
\emph{a priori} obvious: The WCA particles enforce very low density of
oxygen atoms outside a region $-w/2 < z < w/2$; given that water
molecules are not point particles, however, the most realistic
continuum description could involve an offset $\delta$ between $w$ and
$\ell_{\rm w}$, i.e., $\ell_{\rm w} = w-\delta$. Considerations for
choosing $\delta$ will be discussed later.

The two charged planes at $z=\pm w/2$ are treated in our continuum
calculation as uniformly charged sheets with surface charge density
$q\equiv Q/A$, where $A$ is the cross-sectional area of the simulation
cell orthogonal to $z$. Within DCT, these charged planes enter the
continuum model explicitly by introducing a discontinuity of magnitude
$4\pi |q|$ in the total electric field along $z$ (as the planes are
surrounded on either side by vacuum), irrespective of whether they are
coincident with the WCA particles. In contrast, the WCA centers only
enter DCT implicitly by confining the water molecules such that the
thickness of the dielectric slab is $\ell_{\rm w}\equiv w -
\delta$. The continuum representation of the system is summarized in
Fig.~\ref{fig:schematic}b. The simulation cell is periodically
replicated in all three dimensions, and the periodic length along the
$z$-direction is $L$.

In the \tcr{Supplementary Material (SM)}, we solve the periodic
continuum problem shown schematically in Fig.~\ref{fig:schematic}b,
obtaining a total electrostatic potential in the region $-\ell_{\rm
  w}/2 \le z \le \ell_{\rm w}/2$
\begin{widetext}
  \begin{equation}
    \label{eqn:phi_inin}
    \phi(z) = 4\pi q\bigg(-\frac{zw}{L}+z\bigg)
    + 4\pi P\bigg(-\frac{z(w-\delta)}{L}
      + z\bigg),
  \end{equation}
\end{widetext}
where $P$ is the uniform polarization of the dielectric, and we have
assumed that an Ewald-style approach has been used to treat
electrostatic interactions. The first term in Eq.~\ref{eqn:phi_inin}
arises from the charged planes, which we denote $\phi_q$. The second
term arises from the polarized dielectric, and we denote this
$\phi_{\rm solv}$. The total electric field inside the dielectric
follows directly from Eq.~\ref{eqn:phi_inin}:
\begin{equation}
  E = -4\pi q\bigg(1 - \frac{w}{L}\bigg) - 4\pi P\bigg(1 - \frac{w-\delta}{L}\bigg).
\label{eqn:total-field}
\end{equation}
We now combine Eq.~\ref{eqn:total-field} with the local constitutive
relation (Eq.~\ref{eqn:constit}) to obtain an expression for $P$:
\begin{equation}
  \label{eqn:P-LR}
  P = -\frac{(\epsilon-1)(1-\frac{w}{L})q}{1+(\epsilon-1)(1-\frac{w-\delta}{L})}.
\end{equation}

\tcr{We also show in the SM} that the electrostatic potential at the
charged plane at $z=-w/2$ is
\begin{equation}
  \label{eqn:phisolv_lo}
  \phi_{\rm solv, lo} = 2\pi P\bigg(\frac{w(w-\delta)}{L}
    - (w-\delta)\bigg).
\end{equation}
Similarly, for the charged plate at $z=+w/2$ we have
\begin{equation}
  \label{eqn:phisolv_hi}  
  \phi_{\rm solv, hi} = 2\pi P\bigg[-\frac{w(w-\delta)}{L}
    + (w-\delta)\bigg].
\end{equation}
The solvation free energy $f^{(L)}_{\rm solv} = q(\phi_{\rm solv, hi}
- \phi_{\rm solv, lo})/2$ is the difference in reversible work (per
unit area) to introduce the surface charge density $q$ to the charged
planes with and without the solvent present. Combining
Eqs.~\ref{eqn:P-LR},~\ref{eqn:phisolv_lo} and~\ref{eqn:phisolv_hi}
gives
\begin{equation}
  f^{(L)}_{\rm solv}(q) =
  -2\pi q^2(w-\delta)\frac{(\epsilon-1)(1-\frac{w}{L})^2}{1+(\epsilon-1)(1-\frac{w-\delta}{L})}.
\end{equation}
In the limit $L\to\infty$ we recover the expected result
\begin{equation}
  \label{eqn:fsolvinf}
  f^{(\infty)}_{\rm solv}(q) =
  -2\pi q^2\frac{\epsilon-1}{\epsilon}(w-\delta).
\end{equation}
The correction $\Delta f_{\rm DCT}(L) = f^{(\infty)}_{\rm solv} -
f^{(L)}_{\rm solv}$ we should apply for finite $L$ is thus
\begin{equation}
  \label{eqn:fsolvcorr}
  \Delta f_{\rm DCT}(L) = 2\pi
    q^2(w-\delta)(\epsilon-1)\Bigg[\frac{(1-\frac{w}{L})^2}{1+(\epsilon-1)(1-\frac{w-\delta}{L})}
    -\frac{1}{\epsilon}\Bigg].
\end{equation}

Equation~\ref{eqn:fsolvcorr} provides a simple correction term that
can be added to $f^{(L)}_{\rm solv}$ obtained from molecular
simulations. The extent to which $\Delta f_{\rm DCT}(L)$ achieves
consistent estimates of $f^{(\infty)}_{\rm solv}$ from simulations
with different $L$ is then one indicator of how well simple DCT
describes the dielectric properties of water films.

\section{Assessing the accuracy of DCT with molecular simulations}
\label{eqn:ApplyToSim}

\begin{figure}[tb]
  \includegraphics[width=8cm]{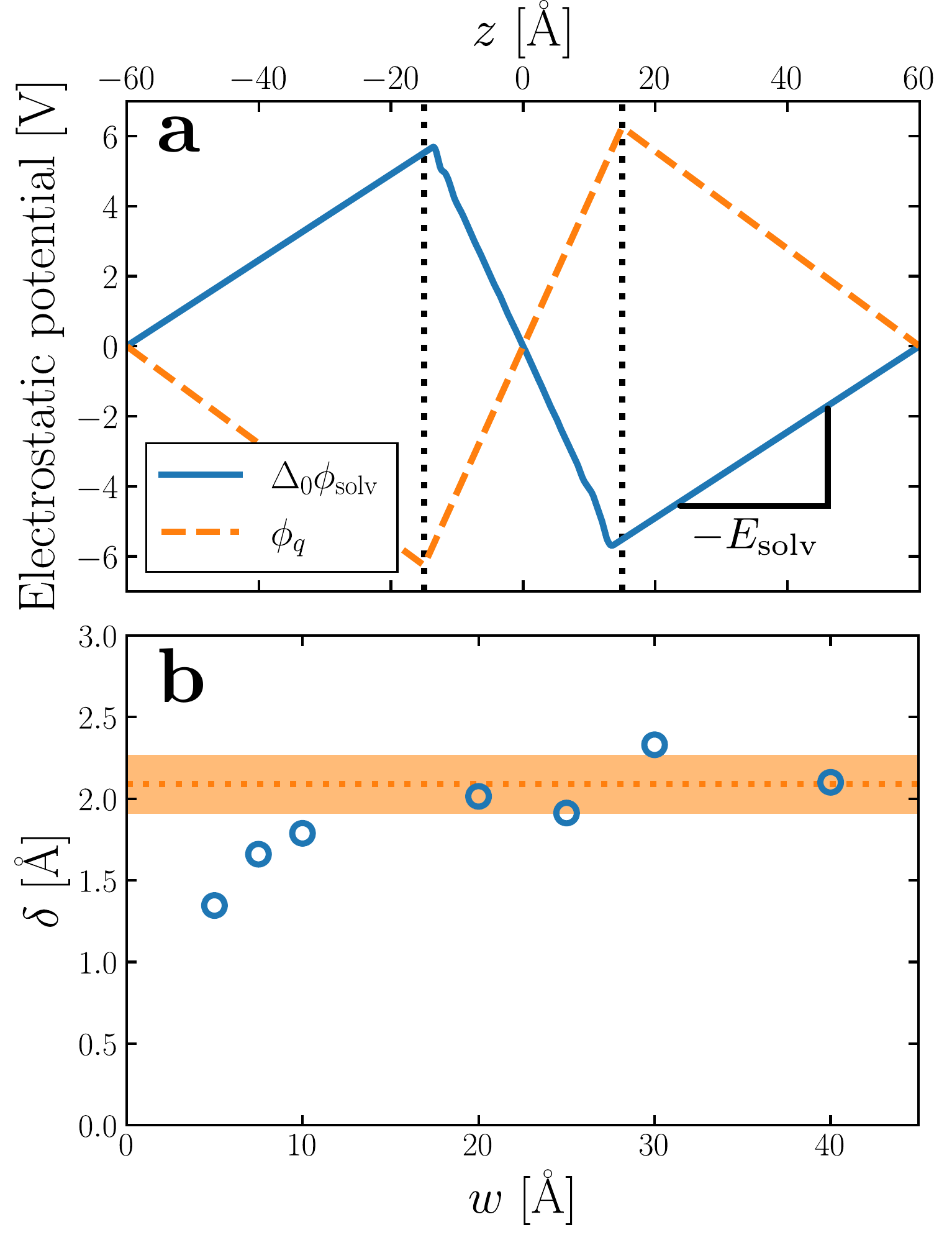}
  \caption{(a) Average electrostatic potential due to the solvent
    (solid blue) and charged planes (dashed orange) with $q\approx
    3\times 10^{-3}$\,$e$/\AA$^2$, $w=30$\,\AA{} and $L=120$\,\AA. The
    vertical dotted lines indicate the positions of the WCA particles.
    For the solvent, we plot $\Delta_0\phi_{\rm solv}(z)=\phi_{\rm
      solv}(z)-\phi_{\rm solv,0}(z)$, where $\phi_{\rm solv,0}$ is the
    average potential with $q=0$\,$e$/\AA$^2$. The average electric
    field due to the solvent is used to determine $\delta$. (b) The
    inferred displacement $\delta/2$ between WCA particles and the
    dielectric boundary depends only weakly on the width of the liquid
    slab.  Each point in the plot of $\delta$ vs. $w$ is the average
    of 5 simulations with different values of $L$. Averaging results
    for $w\ge 20$\,\AA{} gives $\delta = 2.09 \pm 0.17$\,\AA, which is
    used throughout. The shaded orange region indicates a 95\%
    confidence interval. \tcb{Note that in both (a) and (b), results
      have been obtained from simulations where the WCA particles and
      charged planes coincide.}}
  \label{fig:composite_PhiSolvDel}
\end{figure}

To assess our continuum prediction of the finite size correction
$\Delta f_{\rm DCT}(L)$ given by Eq.~\ref{eqn:fsolvcorr}, we will
assume that $\epsilon$ retains its bulk liquid value ($\epsilon_{\rm
  liq}\approx 71$ for SPC/E) over the entire domain $-\ell_{\rm w}/2 <
z < \ell_{\rm w}/2$. The only undetermined parameter in
Eq.~\ref{eqn:fsolvcorr} is then the length scale $\delta$, which
determines the location of the dielectric boundaries of the solvent
relative to the charged planes at $z=\pm w/2$. To determine an
appropriate value of $\delta$, we note that DCT predicts an electric
field due to the solvent in the region $w/2 \le z < L/2$ 
\begin{equation}
  \label{eqn:Esolv-hi}
  E_{\rm solv} = 4\pi P\frac{(w-\delta)}{L}.
\end{equation}
As shown in Fig.~\ref{fig:composite_PhiSolvDel}a, $E_{\rm solv}$ can
be easily obtained from simulation. (Note that, owing to the charge
asymmetric distribution of individual water molecules, $\phi_{\rm
  solv}(z)$ for liquid water varies across the interface even with
$q=0$\,$e$/\AA$^2$. As we are concerned with the \emph{response} of
the dielectric slab, in Fig.~\ref{fig:composite_PhiSolvDel}a we have
plotted $\Delta_0\phi_{\rm solv}(z) \equiv \phi_{\rm solv}(z) -
\phi_{\rm solv,0}(z)$, where $\phi_{\rm solv,0}(z)$ is the average
electric potential profile with $q=0$\,$e$/\AA$^2$.)

For the time being, we consider cases where the charged planes and WCA
centers coincide. For a given $w$, we then measure $E_{\rm solv}$ with
$q\approx 3\times 10^{-3}\,e/{\rm\AA}^2$ for each value of $L$
investigated, and by \tcb{substituting $P$ given by Eq.~\ref{eqn:P-LR}
  into Eq.~\ref{eqn:Esolv-hi}}, we determine $\delta$. Results
obtained with different $w$ \tcb{(see Table~\ref{tab:Nwat})} are shown
in Fig.~\ref{fig:composite_PhiSolvDel}b. Despite some noise, $\delta$
appears to plateau as $w$ increases; averaging results for $w\ge
20$\,\AA, we find $\delta = 2.09 \pm 0.17$\,\AA. \tcb{This procedure
  is similar in spirit and effect to that of
  Ref.~\onlinecite{loche2020universal}, which locates a dielectric
  dividing surface based on the average potential drop across a
  polarized water slab. Our approach does not assign special
  significance to the potential at the confining walls. More
  significantly, we find that $\delta$ decreases for sub-nanometer
  films, in contrast to the increase reported by
  Ref.~\onlinecite{loche2020universal} for water between graphene
  sheets.}

\begin{figure*}[tb]
  \includegraphics[width=16cm]{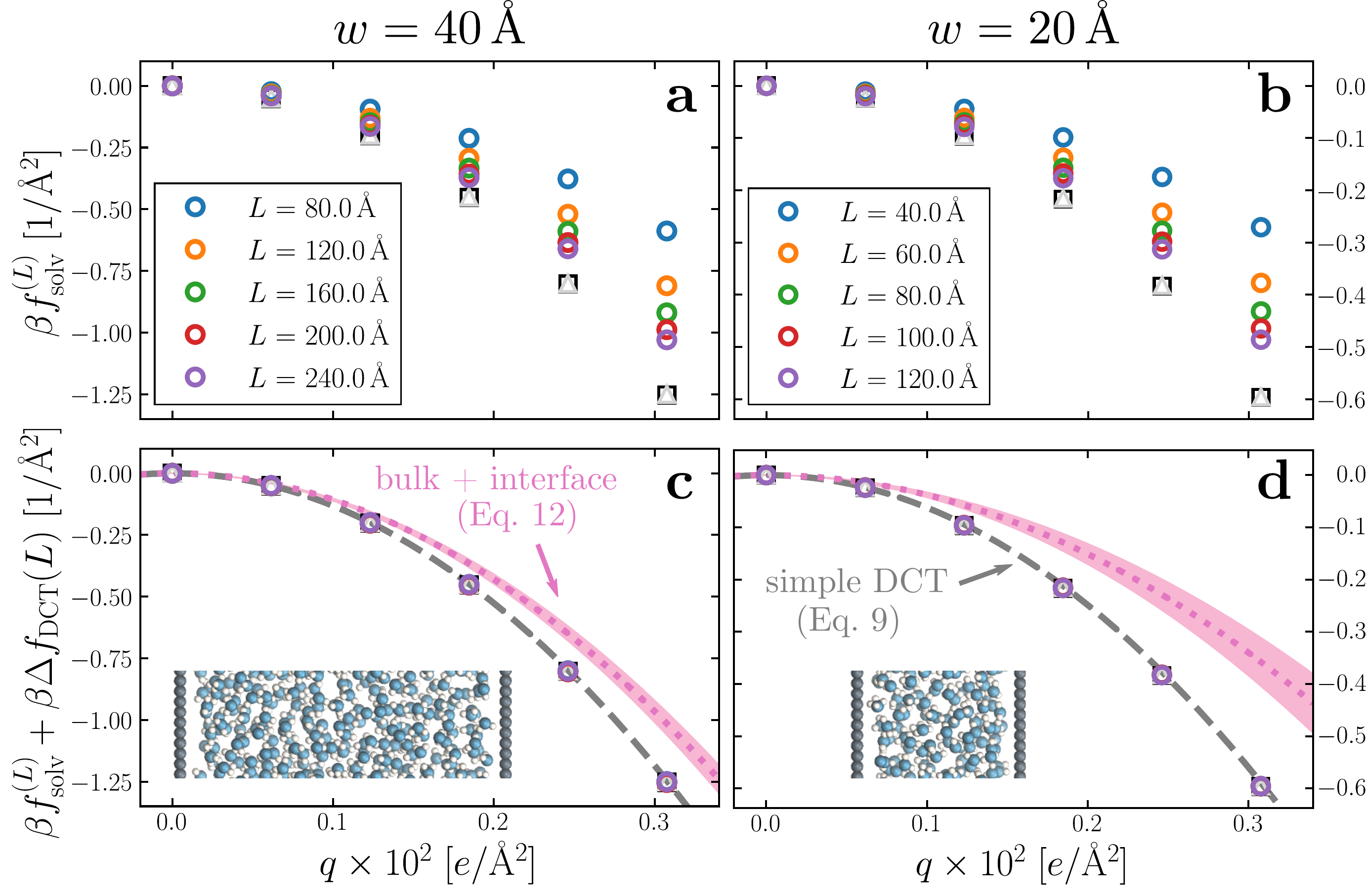}
  \caption{Dependence of solvation free energy $f_{\rm solv}^{(L)}(q)$
    on system size $L$, shown in (a) and (b) for $w=40$\,\AA{} and
    $w=20$\,\AA, respectively.  The values of $L$ for $w=40$\,\AA{}
    are are indicated in the legend of panel (a); those for the
    thinner liquid slab are shown in (b). \tcb{In both cases the WCA
      particles coincide with the charged planes.}  Adding
    \tcb{$\Delta f_{\rm DCT}(L)$} given by Eq.~\ref{eqn:fsolvcorr}
    largely removes this sensitivity, as seen in (c) and (d) for
    $w=40$\,\AA{} and $w=20$\,\AA, respectively.  DCT predictions for
    $f_{\rm solv}^{(\infty)}(q)$ (Eq.~\ref{eqn:fsolvinf}) are plotted
    as dashed gray lines. Black squares and gray triangles show
    results obtained with $D=0$\,V/\AA{} for the smallest and largest
    values of $L$, respectively. The pink dotted lines show
    predictions of $f^{(\infty)}_{\rm solv,int}$ from a dielectric
    continuum model, in which an interfacial layer of width $\ell_{\rm
      int} = 7.5$\,\AA{} is assigned a permittivity $\epsilon_{\rm
      int}=2.1$ much lower than in bulk liquid, computed from
    (Eq.~\ref{eqn:fsolvinf-int}).  The shaded regions bound
    predictions with $6\,{\rm\AA} \le \ell_{\rm int} \le
    9\,{\rm\AA}$. Insets: Snapshots from corresponding molecular
    dynamics simulations.}
  \label{fig:composite_fesolv}
\end{figure*}

Having determined $\delta$, we are now in a position to test the
appropriateness of the finite size correction given by
Eq.~\ref{eqn:fsolvcorr}. To this end, in
Figs.~\ref{fig:composite_fesolv}a and~\ref{fig:composite_fesolv}b, we
show $f_{\rm solv}^{(L)}(q)$ for $w=40$\,\AA{} and $w=20$\,\AA,
respectively. \tcb{We focus on these values of $w$ as they correspond
  to the extremal values investigated that lie in the plateau region
  in Fig.~\ref{fig:composite_PhiSolvDel}b; results for $w=30$\,\AA{}
  and $w=25$\,\AA{} are included in the SM.} As expected, $f_{\rm
  solv}^{(L)}(q)$ exhibits a dependence on system size.  Adding
$\Delta f_{\rm DCT}(L)$ removes this dependence almost entirely, as
seen in Figs.~\ref{fig:composite_fesolv}c
and~\ref{fig:composite_fesolv}d. Also shown are results obtained by
imposing vanishing electric displacement field $D=0$\,V/\AA{} along
$z$, which is formally equivalent to the commonly used Yeh-Berkowitz
approach for approximating 2D Ewald summation
\cite{YehBerkowitz1999sjc,zhang2016finite}. As results obtained with
$D=0$\,V/\AA{} should approximate $L\to\infty$, they do not require a
finite size correction. Importantly, excellent agreement with
$f^{(L)}_{\rm solv}+\Delta f_{\rm DCT}(L)$ is observed, giving us
confidence that Eq.~\ref{eqn:fsolvcorr} provides a meaningful finite
size correction.

The fact that the simple DCT model outlined in
Sec.~\ref{sec:dct-outline} describes the finite size behavior of
$f_{\rm solv}^{(L)}$ so well suggests it is reasonable to think of
thin water films as having a uniform dielectric constant equal to that
of bulk in the region they occupy.  Even more tellingly, the
extrapolation $f_{\rm solv}^{(L)}+\Delta f_{\rm DCT}(L)$ from
simulation agrees well with the continuum prediction $f_{\rm
  solv}^{(\infty)}$ in Eq.~\ref{eqn:fsolvinf}.

\begin{figure}[tb]
  \includegraphics[width=8cm]{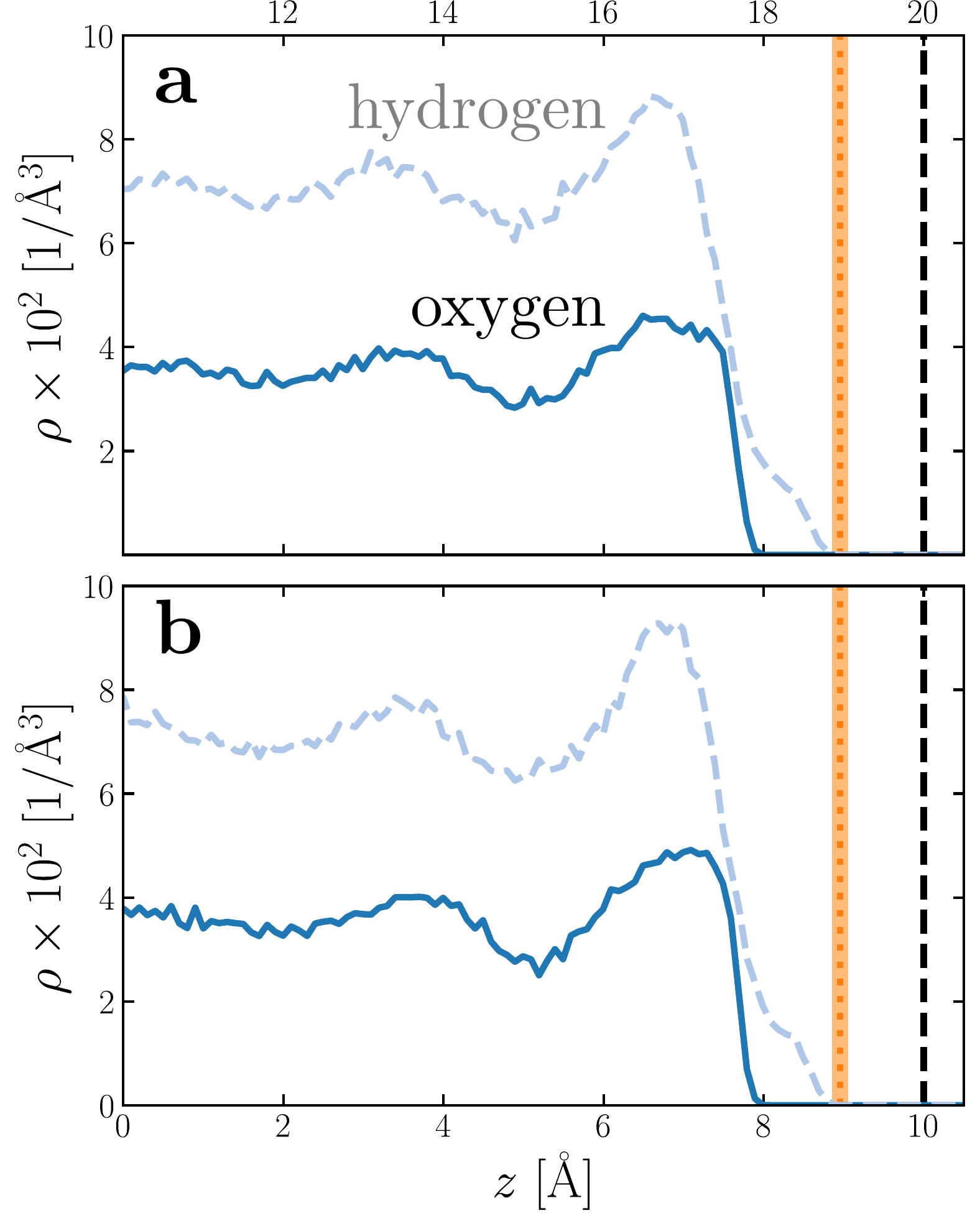}
  \caption{Number density profiles $\rho(z)$ for hydrogen (dashed
    blue) and oxygen (solid blue) atoms of water, with
    $q=0$\,$e$/\AA$^2$ for (a) $w=40$\,\AA{} and (b)
    $w=20$\,\AA. \tcb{In both cases the WCA particles coincide with
      the charged planes.} The vertical dashed line shows the
    location $z=w/2$ of WCA particles, and the vertical dotted line
    indicates the dielectric boundary at $z=(w-\delta)/2$. (The shaded
    region indicates the same 95\,\% confidence interval as in
    Fig.~\ref{fig:composite_PhiSolvDel}.) In both cases, the
    dielectric boundary aligns closely with the vanishing of hydrogen
    atom density.}
  \label{fig:composite_DensDel}
\end{figure}

To provide a physical interpretation for the length scale $\delta$,
Fig.~\ref{fig:composite_DensDel} shows number density profiles
$\rho(z)$ for water's oxygen and hydrogen atoms from simulations with
$q=0$\,$e$/\AA$^{2}$. On these plots, we have also marked the boundary
predicted by $\ell_{\rm w}/2 = (w-\delta)/2$, which corresponds
closely to the vanishing of average hydrogen density. Because the
hydrogen atoms protrude further toward the vapor phase than the oxygen
atoms, this boundary marks the outermost limit of microscopic sources
of polarization fluctuations. The water film thickness we have
inferred is thus the largest that could be reasonably justified based
on the statistics of molecular configurations.

\section{Assessing the validity of other models}
\label{sec:OtherModels}

We have shown that $f_{\rm solv}^{(L)}+\Delta f_{\rm DCT}(L)$ obtained
from simulation agrees well with the predictions of a simple DCT in
which the dielectric constant of thin films is identical to that of
the bulk liquid (Eq.~\ref{eqn:fsolvinf}). If we were to decrease
$\epsilon$, agreement with simulation data would require assigning
$\ell_{\rm w}$ a larger value than we have inferred, i.e., a value
that would be difficult to justify from microscopic structure. This
observation advocates against the notion that the overall dielectric
permittivity of the thin film is lower than in the homogeneous
fluid. By itself, however, it does not rule out a model in which the
interfacial regions have a permittivity $\epsilon_{\rm int}$ that is
distinct from the bulk region they sandwich. For such a model, the
free energy reads
\begin{equation}
    \label{eqn:fsolvinf-int}
  f^{(\infty)}_{\rm solv,int} =
  -2\pi q^2\bigg[\ell_{\rm bulk}\left(\frac{\epsilon-1}{\epsilon}\right)
    + 2\ell_{\rm int}\left(\frac{\epsilon_{\rm int}-1}{\epsilon_{\rm int}}\right)\bigg].
\end{equation}
where $\ell_{\rm int}$ is the width of each interfacial region.

Following the dielectric imaging experiments of Fumagalli \etal
\cite{fumagalli2018anomalously}, we take $\ell_{\rm int} = 7.5\pm
1.5$\,\AA{} and $\epsilon_{\rm int} = 2.1$, and require the total
width $\ell_{\rm w} = \ell_{\rm bulk}+2\ell_{\rm int}$ to have the
same value as in the uniform dielectric model: As discussed above, it
is unreasonable to allow $\ell_{\rm w}$ to increase from that value.
Decreasing $\ell_{\rm w}$, on the other hand, offers less flexibility
to a model that introduces regions of low dielectric constant at the
expense of those with high dielectric constant. The resulting
predictions of $f^{(\infty)}_{\rm solv,int}$ are shown in
Figs.~\ref{fig:composite_fesolv}c and~\ref{fig:composite_fesolv}d
(labeled ``bulk+interface''), where poor agreement with the simulation
data is observed.  Quantitatively different (but not significantly
improved) predictions would be obtained with different choices of
$\ell_{\rm int}$ and $\epsilon_{\rm int}$. We find generally that
$\ell_{\rm int}=0$ (or equivalently, $\epsilon_{\rm int}=\epsilon_{\rm
  liq}$) yields the best agreement with simulation. Evidence for this
conclusion is provided in SM.

\begin{figure}[b]
  \includegraphics[width=8cm]{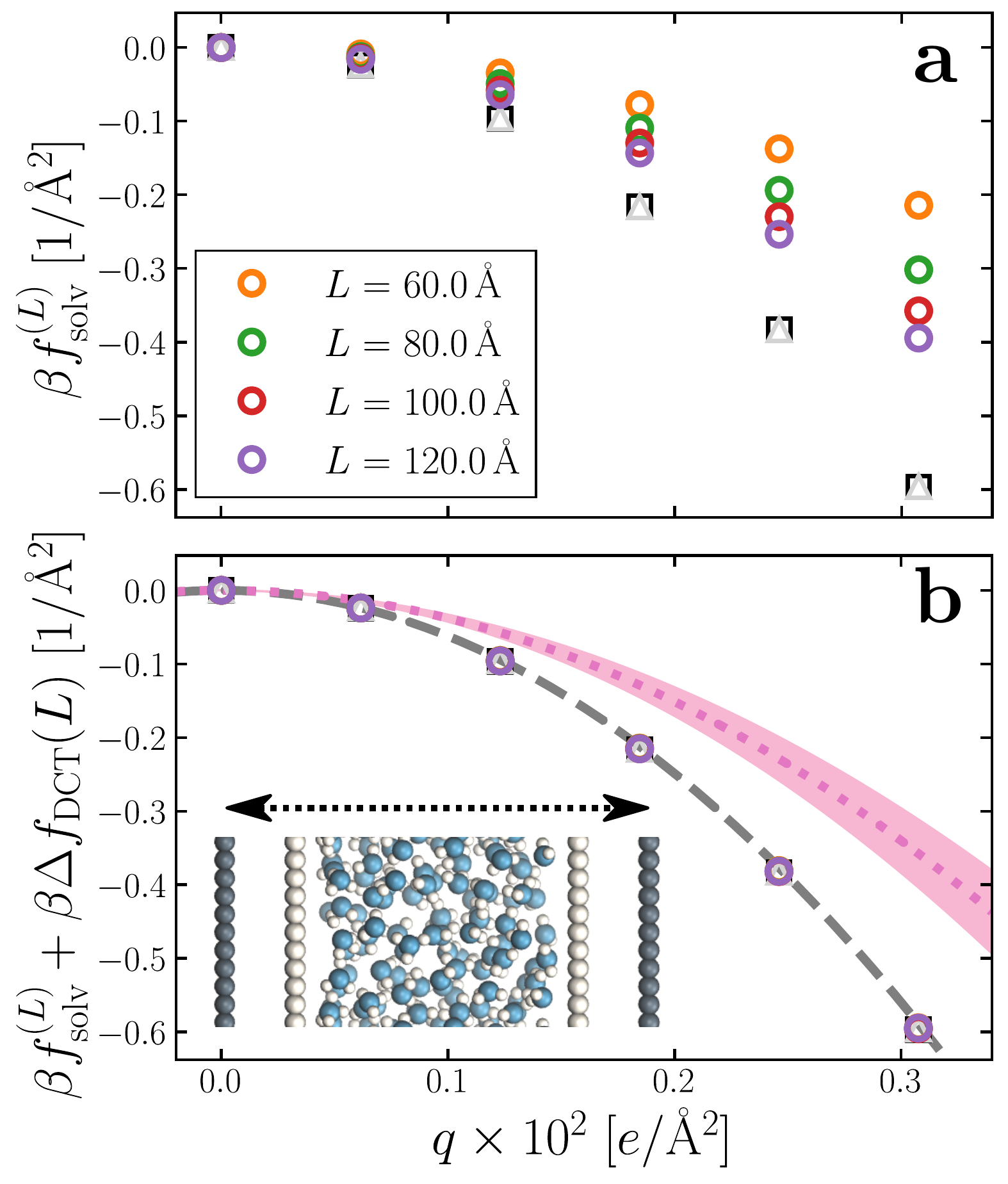}
  \caption{Solvation free energies with the charged planes moved
    5\,\AA{} into vacuum. (a) $f_{\rm solv}^{(L)}(q)$ exhibits the
    same finite size dependence as in
    Fig.~\ref{fig:composite_fesolv}b, implying the same value of
    $w-\delta$ and thus demonstrating that the layer of width
    $\delta/2$ should not be associated with the liquid phase.  (b)
    Adding \tcb{$\Delta f_{\rm DCT}(L)$} to the results in (a), with
    $w=30$\,\AA{} and $\delta=12.09$\,\AA, essentially removes
    dependence on $L$ entirely.  Inset: Snapshot from a molecular
    dynamics simulation showing the position of the charged planes
    relative to the WCA centers (see Fig.~\ref{fig:schematic}). The
    double headed arrow indicates $w$.}
  \label{fig:composite_fesolv_displ}
\end{figure}

The width $\ell_{\rm int}$ of a notional interfacial layer differs
fundamentally from the length scale $\delta$ in our simple uniform
DCT. They can nonetheless easily be confused. In our case $z = \pm
(w-\delta)/2$ marks the location of a sharp interface between vapor
and bulk liquid. This interface does not coincide with the location
$z=\pm w/2$ of the confining charged walls because their constituent
WCA particles exclude volume. $\delta$ thus characterizes a region
that is inaccessible to water molecules and should not be associated
with the liquid. To emphasize this point, we modify the $w=20$\,\AA{}
system (Figs.~\ref{fig:composite_fesolv}b
and~\ref{fig:composite_fesolv}d) by displacing the charged planes
5\,\AA{} into vacuum, with the WCA particles fixed at their original
positions (i.e., the general case considered in
Fig.~\ref{fig:schematic}a).  $\delta$ increases by 5\,\AA{} as a
result, while $\ell_{\rm w} = w-\delta$ is unchanged, i.e., $w\to
30$\,\AA{} and $\delta\to 12.09$\,\AA{}, while $\ell_{\rm w} =
17.91$\,\AA{} just as before. Changing $\delta$ in this fashion
clearly has nothing to do with water's interfacial dielectric
properties.  Fig.~\ref{fig:composite_fesolv_displ} presents results
for $f_{\rm solv}^{(L)}$ and $f_{\rm solv}^{(L)}+\Delta f_{\rm
  DCT}(L)$ for the displaced-charge system, which are virtually
indistinguishable from their undisplaced counterparts in
Fig.~\ref{fig:composite_fesolv}.

By contrast, a layer of width $\ell_{\rm int}$ in ``bulk+interface''
models is clearly associated with the liquid. It is imagined to
comprise water molecules whose orientational fluctuations are distinct
from those in bulk liquid due to the phase boundary.  Multiple studies
based on such models have concluded that the interfacial layer has a
greatly reduced polarizability, amounting to a ``dead layer'' with
$\epsilon_{\rm int}\approx 1$
\cite{zhang2018note,zhang2018note,matyushov2021dielectric,mondal2021anomalous,fumagalli2018anomalously}. Dielectric
properties of this notional dead layer may be nearly indistinguishable
from vacuum, but the layer plainly belongs to the dense liquid phase
within a ``bulk+interface'' picture.

\section{Ultra thin films of water}
\label{sec:SmallW}

\begin{figure}[tb]
  \includegraphics[width=8cm]{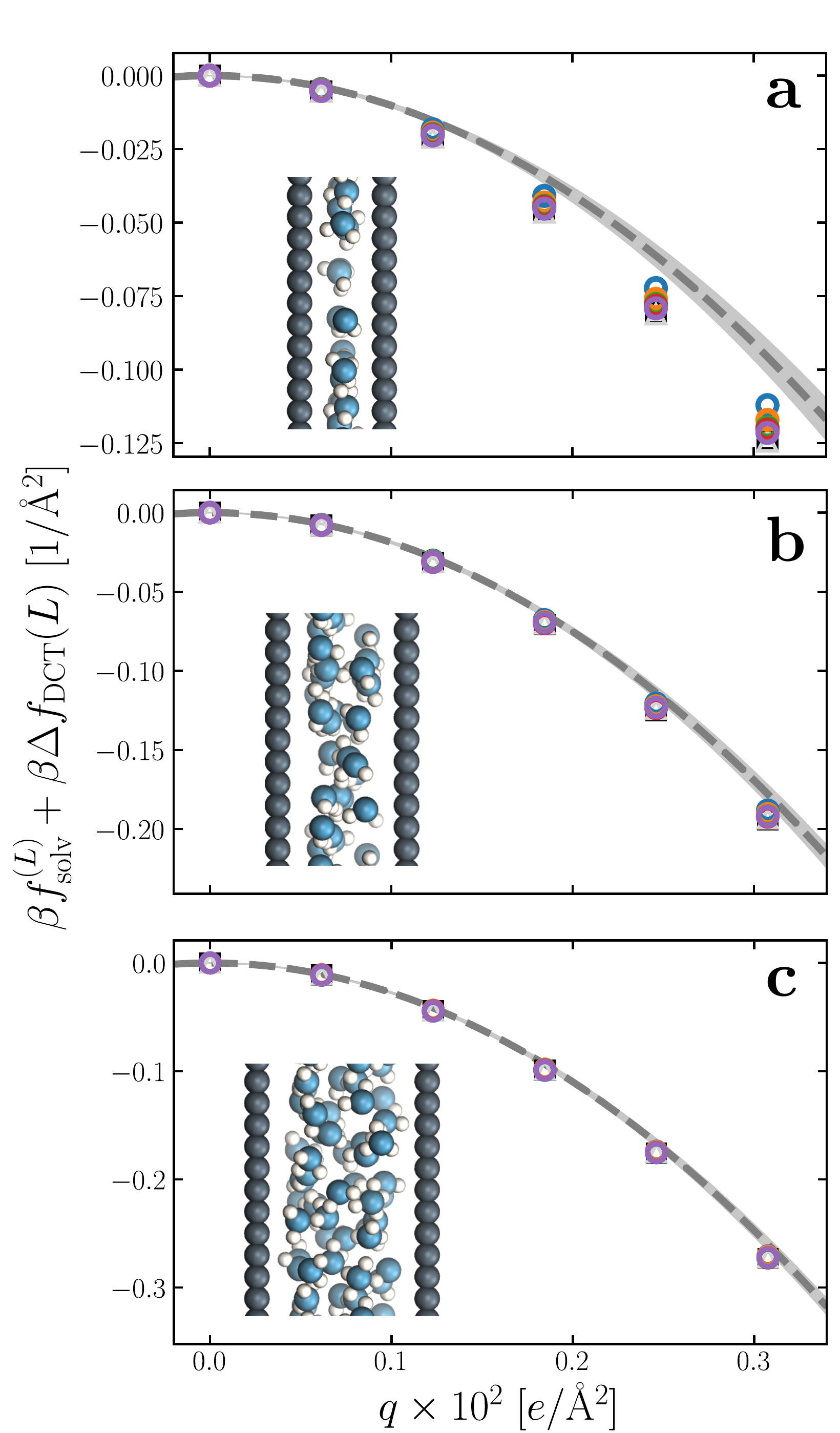}
  \caption{Solvation free energy \tcb{$f_{\rm solv}^{(L)}(q) + \Delta
      f_{\rm DCT}(L)$} in ultra thin water films, for (a) $w=5$\,\AA,
    (b) $w=7.5$\,\AA, and (c) $w=10$\,\AA. For each value of $w$,
    simulations with $L=2w,3w,\ldots,6w$ have been performed. \tcb{In
      all cases the WCA particles coincide with the charged planes.}
    The dashed line shows $f_{\rm solv}^{(\infty)}(q)$ predicted by
    DCT (Eq.~\ref{eqn:fsolvinf}), and the shaded region encompasses
    predictions with $\delta = 2.09\pm 0.17$\,\AA.  Insets: snapshots
    from molecular dynamics simulations.}
  \label{fig:composite_SmallW}
\end{figure}

We have established so far that films of water with $\ell_{\rm w}
\gtrsim 18$\,\AA{} behave quantitavely like simple dielectric continua
with regard to their response to a uniform electric field. We now
investigate the behavior of `ultra thin' films confined between
charged plates with $w\le 10$\,\AA. In
Figs.~\ref{fig:composite_SmallW}a, \ref{fig:composite_SmallW}b and
\ref{fig:composite_SmallW}c, we show $f^{(\infty)}_{\rm solv}$
obtained with simulation for $w=5$\,\AA, 7.5\,\AA, and 10\,\AA,
respectively, using $\delta = 2.09\pm 0.17$\,\AA{} to correct for
finite size effects (Eq.~\ref{eqn:fsolvcorr}). As the thickness of the
water slab is reduced to length scales comparable to
$\ell_\epsilon\approx 6$\,\AA, treating the water molecules as a
dielectric continuum is certainly questionable. Discrepancies between
$f_{\rm solv}^{(\infty)}$ (Eq.~\ref{eqn:fsolvinf}) and $f_{\rm
  solv}^{(L)}+\Delta f_{\rm DCT}(L)$ indeed become apparent as $w$ is
decreased below 1\,nm, but the relative error of continuum predictions
is surprisingly modest. Even when $w$ is only large enough to
accommodate a single molecular layer
(Fig.~\ref{fig:composite_SmallW}a), the continuum prediction in
Eq.~\ref{eqn:fsolvinf} provides a reasonable ballpark estimate of the
solvation free energy. For two to three molecular layers
(Fig.~\ref{fig:composite_SmallW}b and~\ref{fig:composite_SmallW}c),
quantitative agreement between simple DCT and the simulation data is
recovered almost entirely.

\section{Reconciling our results with dielectric imaging experiments}
\label{sec:ExpComp}

\begin{figure}[tb]
  \includegraphics[width=8cm]{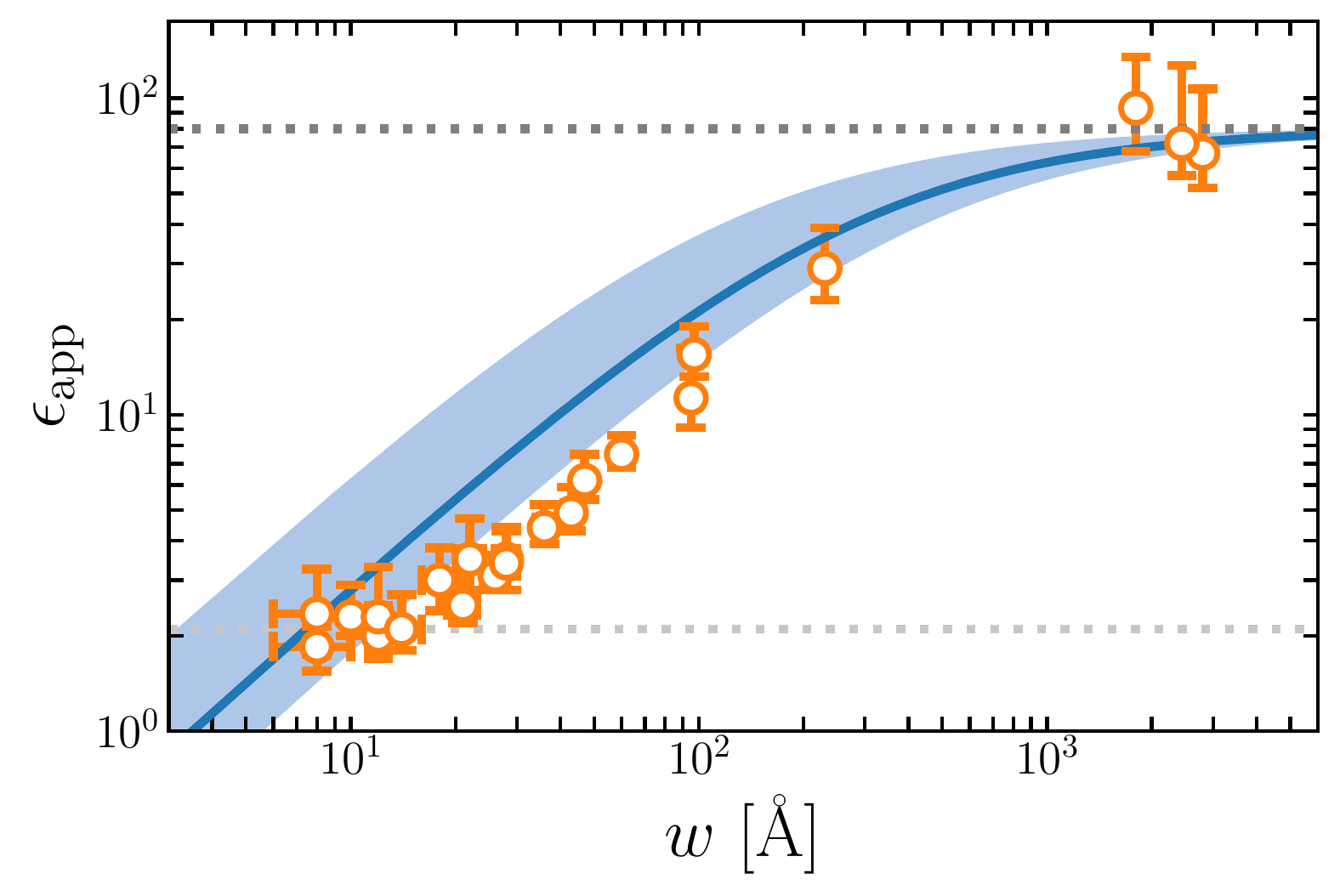}
  \caption{The apparent dielectric constant $\epsilon_{\rm app}$
    predicted by simple DCT (Eq.~\ref{eqn:epsprime}) is broadly
    consistent with experiment. The blue solid line is obtained with
    $\delta = 3.5$\,\AA, which is estimated from \emph{ab initio}
    molecular dynamics simulations of water on graphene. The blue
    shaded region indicates the range of $\epsilon_{\rm app}$ obtained
    with $\delta = 3.5\pm 2.0$\,\AA, demonstrating the sensitivity of
    $\epsilon_{\rm app}$ to uncertainty in the film thickness. The
    light and dark gray dotted lines indicate $\epsilon_{\rm app} =
    2.1$ and $\epsilon_{\rm app} = 80$, respectively.}
  \label{fig:ExpComp}
\end{figure}

The conclusion we have drawn from computer simulations---that the
dielectric response of nanoscopically thin water films can be
anticipated from bulk properties alone---is squarely at odds with the
conclusion drawn by Fumagalli \etal{}\cite{fumagalli2018anomalously}
based on dielectric imaging measurements of confined water. In this
section we attempt to reconcile our results with those
measurements. Assuming that our simple uniform continuum model is
correct, we show how uncertainty in the thickness of a water film can
cause the apparent dielectric constant $\epsilon_{\rm app}$ to depend
sensitively on film thickness. More specifically, we assess the
consequences of assigning a width $l_{\rm w}+\delta$ to a film whose
actual thickness is $l_{\rm w}$, i.e., failing to account for the
volume excluded by a confining substrate. Based on this assignment and
the development in Sec.~\ref{sec:dct-outline}, we would expect a
solvation free energy
\begin{equation}
  \label{eqn:fsolvinf-wonly}
  f_{\rm solv, app}^{(\infty)}(q) = -2\pi q^2 \frac{\epsilon_{\rm app}
    -1}{\epsilon_{\rm app}}w.
\end{equation}
Equating this with the ``true'' free energy in Eq.~\ref{eqn:fsolvinf},
we obtain
\begin{equation}
  \label{eqn:epsprime}
  \epsilon_{\rm app}(w) = \left(1-\frac{\epsilon-1}{\epsilon}\frac{w-\delta}{w}\right)^{-1},
\end{equation}
which depends explicitly on the film's thickness. While similar
functional forms for $\epsilon_{\rm app}(w)$ have been reported
previously \cite{zhang2018note,matyushov2021dielectric}, the physical
interpretation here is different: As discussed in
Sec.~\ref{sec:OtherModels}, $\delta$ is not to be associated with the
properties of interfacial water.

Fig.~\ref{fig:ExpComp} plots the apparent permittivity $\epsilon_{\rm
  app}$ in Eq.~\ref{eqn:epsprime} as a function of $w$.  Here we have
set $\epsilon = \epsilon_{\rm liq}=80$ and estimated $\delta \approx
3.5$\,\AA{} for the graphene-water interface (based on density
profiles obtained from \emph{ab initio} molecular dynamics simulations
\cite{tocci2020ab}).  Diminished values of $\epsilon_{\rm app}$ at
small $w$ could easily, {\em but mistakenly}, be taken to signify a
strong suppression of polarization fluctuations and response in
nanoscale water films.

The inference of suppressed interfacial permittivity from experiments
may suffer from the same issues that cause $\epsilon_{\rm app}$ to
depend strongly on film thickness\tcb{, much as suggested by
  Ref.~\onlinecite{loche2020universal}.}  To emphasize this point, in
Fig.~\ref{fig:ExpComp} we include dielectric imaging data from
Ref.~\onlinecite{fumagalli2018anomalously}, which exhibit a very
similar dependence on $w$. As an important caveat, the samples studied
by Fumagalli \etal{} have a more complicated geometry than the simple
``slit-pore'' scenario we have considered, involving an AFM tip,
multiple water channels, a graphite substrate, and hexagonal boron
nitride walls. Since geometry of the dielectric boundary is precisely
the issue under scrutiny here, the comparison between theory and
experiment suggested by Fig.~\ref{fig:ExpComp} should be made
cautiously and only qualitatively.  In our view it nonetheless
suggests that the correct interpretation of measurements in
Ref.~\onlinecite{fumagalli2018anomalously} may not in fact require
invoking an interfacial dead layer.

\section{Summary}
\label{sec:Summary}

In this article, we have probed the dielectric response of thin water
films using molecular simulations with finite size corrections from
DCT. We specifically calculated the solvent contribution to the
reversible work of introducing charge to parallel plates that confine
the film. Our results demonstrate that response to such slowly varying
external fields can be accurately captured with a DCT model whose
permittivity is simply equal to that of the homogeneous liquid, in the
entire region occupied by the liquid. Our analysis reveals that
appropriate dielectric boundaries for these films extend to the point
where the hydrogen number density approximately vanishes, and thus
incorporate all microscopic sources of polarization fluctuations. This
observation is consistent with our recent study, where we found that
the dielectric boundary between water and spherical solutes was
reasonably described by the first peak in the solute-hydrogen radial
distribution function \cite{cox2021quadrupole}. Within this simple DCT
approach, which achieves quantitative agreement with simulation for
films $\gtrsim 1$\,nm, water's interfacial regions do not enter as
separate domains. \tcb{Loche \emph{et al.} \cite{loche2020universal}
  have similarly concluded that water films a few nanometers in
  thickness are well characterized by bulk dielectric parameters, but
  they report substantial deviations at smaller scales.}

We also demonstrated rough consistency with experiments
\cite{fumagalli2018anomalously} that had previously been interpreted
to imply a dielectrically dead layer of water at the liquid's
boundary.  This agreement is achieved by asserting that dielectric
boundaries had previously not been placed appropriately. \tcb{For the
  simple point charge model used in this study, the point where the
  hydrogen number density approximately vanishes coincides with the
  point where microscopic charge density vanishes. For polarizable
  models or \emph{ab initio} treatments of water, it is possible that
  the distribution of electron density beyond the hydrogen atoms also
  plays a role \cite{kathmann2011understanding}. Further investigation
  of this point is left for future work.}

Our conclusion is further supported by results for subnanometer water
films that comprise one to three molecular layers. In these cases, one
cannot sensibly discuss a bulk region, yet the simple DCT model still
performs remarkably well.  If anything, the apparent dielectric
constant would need to \emph{increase} to improve agreement with the
free energy data. \tcb{This result contrasts with conclusions of
  Refs.~\onlinecite{schlaich2016water,loche2020universal}, which
  suggest greatly suppressed polarizability at comparable scales of
  confinement.}

To be certain, applying simple DCT to these subnanometer films cannot
be carefully justified (see
Sec.~\ref{sec:OverviewDielectrics}). Nonetheless, this \emph{ad hoc}
application of simple DCT to small length scales strongly argues
against the notion of an interfacial region with low dielectric
constant. Our conclusion is also in line with previous studies that
have found corrections similar to \tcb{$\Delta f_{\rm DCT}(L)$} for
the solvation of small spherical ions in water work remarkably well
down to the nanometer scale, for both bulk
\cite{HummerGarcia1996sjc,FigueiridoLevy1995sjc,HunenbergerMcCammon1999sjc}
and interfacial systems \cite{cox2018interfacial}. Similarly, the
simple DCT model described in this article has also been found to
accurately capture mean field-like corrections for thin films of water
where electrostatic interactions are treated in a short-ranged
fashion\cite{cox2020dielectric}.

\tcb{The approach we have taken is not well suited to address a free
  liquid-vapor interface, whose substantial topographical fluctuations
  greatly complicate formulating and solving an appropriate dielectric
  boundary value problem. Previous work on the free interface has
  emphasized that in representative configurations the liquid phase
  terminates sharply at any given lateral position
  \cite{WillardChandler2010sjc,willard2014molecular}.  Since our
  results stress the importance of precisely establishing the liquid's
  microscopic boundary, we expect that dielectric models based on a
  smoothed average interface are a poor caricature of this scenario.
  Instead, a faithful assessment of permittivity at the free
  liquid-vapor interface will require attention to its undulating
  instantaneous structure, suggesting that a more spatially localized
  analysis is likely more feasible.}

It would be incorrect, however, to conclude from our results that
simple DCT gives a full account of polarization response at the
liquid's surface. Indeed, even in bulk liquid water, the charging of
small spherical solutes is characterized by short-wavelength solvent
response that is not well-described by simple continuum approaches
\cite{HummerGarcia1996sjc}. In such cases, and in contrast to the thin
films considered in this work, $f^{(\infty)}_{\rm solv}$ predicted by
DCT is a poor estimate of that obtained from simulations
\cite{cox2018interfacial,cox2021quadrupole}. The impact of such
profound perturbations are even more pronounced for solutes near soft
interfaces like that between water and its vapor, where distortions of
the interface result in nonlinearities beyond the scope of current
theoretical treatments \cite{cox2020assessing}. But for perturbations
that vary slowly in space, like the uniform fields considered here,
the results of this study add to a growing body of work that supports
a surprisingly simple view of water's surface (and, by extension, thin
films) as a dielectric medium: Its local permittivity is equivalent to
the bulk dielectric constant, all the way down to nanometer length
scales.

\section{Methods}
\label{sec:Methods}

All simulations followed the basic setup shown in
Fig.~\ref{fig:schematic}a. Two planes of $N_{\rm site} = 100$ point
charges were placed on a square lattice at $z=\pm w/2$. Water
molecules, modeled with the SPC/E potential
\cite{BerendsenStraatsma1987sjc}, were confined to the region $-w/2\le
z\le w/2$ by volume-excluding WCA particles \cite{WCA}, whose centers,
for the most part, coincided with the point charges at $z=\pm
w/2$. The interaction between an individual WCA particle and a water
molecule is defined by
\begin{equation}
  u_{\rm WCA}(r) =
  \begin{cases}
    4\varepsilon\left[\left(\frac{\sigma}{r}\right)^{12}-\left(\frac{\sigma}{r}\right)^6\right], & r\le 2^{1/6}\sigma, \\
    0, & r>2^{1/6}\sigma,
  \end{cases}
\end{equation}
where $\sigma = 2.5$\,\AA, $\varepsilon = 0.1$\,kcal/mol, and $r$ is
the distance between the WCA particle and the oxygen atom of the water
molecule. As described in Sec.~\ref{sec:OtherModels}, for $w=20$\,\AA,
we also performed simulations where the planes of point charges were
displaced 5\,\AA{} into vacuum, but leaving the rest of the system
unchanged. For each value of $w$, simulations of 5\,ns (following at
least 100\,ps of equilibration) were performed with $q_{\rm site}/e = 0,
1\times 10^{-3},\ldots,5\times 10^{-3}$. The total volume of the
simulation cells was $12.75\,{\rm \AA}\times 12.75\,{\rm \AA}\times
L$, where $L$ takes values as indicated throughout the
article.

All simulations were performed with the \texttt{LAMMPS} simulations
package \cite{plimpton1995sjc}. Lennard-Jones interactions between
water molecules were truncated and shifted at 10\,\AA. Long range
electrostatic interactions were evaluated using particle-particle
particle-mesh Ewald summation \cite{HockneyEastwood1988sjc}, with
parameters chosen such that the RMS error in the forces were a factor
$10^{5}$ smaller than the force between two unit charges separated by
a distance of 1.0\,\AA \cite{kolafa1992cutoff}. Where indicated in the
text, the electric displacement field along $z$ was set to zero, using
the implementation given in Ref.~\onlinecite{cox2019finite}. The
geometry of the water molecules was constrained using the
\texttt{RATTLE} algorithm \cite{andersen1983rattle}. Temperature was
maintained at 298\,K with a Nos\'{e}-Hoover chain thermostat
\cite{shinoda2004rapid,tuckerman2006liouville} with a damping constant
of 100\,fs. A time step of 2\,fs was used throughout.  The number of
water molecules used in the simulations is given in
Table~\ref{tab:Nwat}.

\begin{table}[h!]
  \centering
  \caption{Number of water molecules $N_{\rm wat}$ for each value of
    $w$ investigated (WCA centers coincide with point charges).}
  \label{tab:Nwat}
  \begin{tabular}{c c c c c c c c}
    \hline
    \hline
    $w$/\AA{}   & 5  & 7.5 & 10 & 20 & 25  & 30  & 40  \\
    \hline
    $N_{\rm wat}$ & 14 & 27  & 41 & 93 & 125 & 143 & 206 \\
    \hline
  \end{tabular}
\end{table}

The free energy of charging up parallel plate capacitors was computed
by averaging electric potentials appropriately.  Let
$\varphi^{(i)}_{\rm solv,hi}(\mbf{R}^N)$ and $\varphi^{(j)}_{\rm
  solv,lo}(\mbf{R}^N)$ denote the instantaneous electric potentials
due to the solvent with configuration $\mbf{R}^{N}$ at site $i$ of one
of the point charges in the plane at $z=w/2$, and site $j$ in the
plane at $z=-w/2$, respectively. The total solvation free energy
$F_{\rm solv}^{(L)}(Q)$ is then defined by
\begin{align}
  &\exp\left[-\beta F_{\rm solv}(Q)\right] = \langle\exp\left[-\beta Q\Delta\varphi_{\rm solv}\right]\rangle_0^{(L)} \nonumber \\[7pt]
  &= \int\!\mrm{d}(\Delta\varphi_{\rm solv})\,P^{(L)}(\Delta\varphi_{\rm solv};0)\exp\left[-\beta Q\Delta\varphi_{\rm solv}\right],
\end{align}
where
\begin{align}
  \Delta\varphi_{\rm solv}(\mbf{R}^{N}) &=
  \frac{1}{N_{\rm site}}\left(
  \sum_{i\in\rm hi}\varphi^{(i)}_{\rm solv,hi}(\mbf{R}^N) -
  \sum_{j\in\rm lo}\varphi^{(j)}_{\rm solv,lo}(\mbf{R}^N)\right) \nonumber \\
  &= \varphi_{\rm solv,hi}(\mbf{R}^N) - \varphi_{\rm solv,lo}(\mbf{R}^N);
\end{align}
$P^{(L)}(\Delta\varphi_{\rm solv};Q)$ is the probability distribution
of $\Delta\varphi_{\rm solv}$ in the presence of two charged planes
with total charges $\pm Q$, in a simulation box of length $L$; and
$\langle\cdot\rangle_0^{(L)}$ denotes an average over
$P^{(L)}(\Delta\varphi_{\rm solv};0)$. 

Similar to our previous
studies\cite{cox2018interfacial,cox2021quadrupole}, $F_{\rm solv}(Q)$
was computed using the \texttt{MBAR}
algorithm\cite{ShirtsChodera2008sjc}. The solvation free energies per
unit area, $f_{\rm solv}^{(L)}$ that we consider are then obtained by
dividing $F_{\rm solv}(Q)$ by the cross-sectional area of the
simulation cell.

\section*{Supplementary Material}
Supplementary Material includes a detailed derivation of results
obtained for the periodic continuum model presented in
Sec.~\ref{sec:dct-outline}. Results for the ``bulk+interface'' model
with different parameters are also presented\tcb{, along with those
  obtained with $w=30$\,\AA{} and $w=25$\,\AA.}

\section*{Author contributions}
\tcb{S.J.C and P.L.G conceived and conducted research. Both authors
  wrote the paper, at all stages.}

\section*{Conflicts of Interest}
\tcb{There are no conflicts to declare.}

\begin{acknowledgments}
We are grateful to Laura Fumagalli for sharing her experimental
results. We also thank Gabriele Tocci for sharing his results from
\emph{ab initio} molecular dynamics. S.J.C is a Royal Society
University Research Fellow (URF\textbackslash R1\textbackslash 211144)
at the University of Cambridge. P.L.G is supported by the
U.S. Department of Energy, Office of Basic Energy Sciences, through
the Chemical Sciences Division (CSD) of Lawrence Berkeley National
Laboratory (LBNL), under Contract DE-AC02-05CH11231. \tcb{For the
  purposes of open access, the authors have applied a Creative Commons
  Attribution (CC BY) licence to any Author Accepted Manuscript
  version arising.}
\end{acknowledgments}

\section*{Data Availability Statement}

The data that supports the findings of this study and input files for
the simulations are openly available at the University of Cambridge
Data Repository, \url{https://doi.org/10.17863/CAM.81959}.

\bibliography{../cox}

\clearpage
\onecolumngrid
\renewcommand\thefigure{S\arabic{figure}}
\renewcommand\theequation{S\arabic{equation}}
\renewcommand\thesection{S\arabic{section}}
\setcounter{figure}{0}
\setcounter{equation}{0}
\setcounter{section}{0}

\noindent {\Large \tbf{Supplementary Material}}

\suppressfloats

\vspace{0.25cm}

\section{Derivation}

The system we consider is shown schematically in
Fig.~\ref{fig:schematic}. Two planes with charge density $\pm q$ are
located at $z=\pm w/2$. We will consider a more general case than in
the main article.  Here, a linear dielectric occupies the region
$-(w/2 - \delta_{\rm lo}) \le z \le (w/2-\delta_{\rm hi})$, such that
$\delta_{\rm lo}+\delta_{\rm hi}=\delta$; while the electrostatic
potential is sensitive to the values of $\delta_{\rm lo}$ and
$\delta_{\rm hi}$, we will show that $\Delta f_{\rm DCT}(L)$ only
depends on their sum. The boundaries of the dielectric are situated at
$\xi_{\rm hi} = w/2 - \delta_{\rm hi}$ and $\xi_{\rm lo} = w/2 -
\delta_{\rm lo}$. The polarization of the medium is $P$.

\subsection*{Potential due to the charged plates}

The potential due to the charged plates is,
\begin{equation}
  \phi_q(z) = 4\pi\int_{\rm cell}\!\mrm{d}z^\prime\rho_q(z^\prime)J(z-z^\prime),
\end{equation}
with
\begin{equation}
  \rho_{q}(z) = q\big[\delta_{\rm D}(z-w/2) - \delta_{\rm D}(z+w/2)\big],
\end{equation}
where $\delta_{\rm D}(x)$ is the Dirac delta-function,
and\cite{wirnsberger2016non,pan2017effect,cox2018interfacial}
\begin{equation}
  J(z) = \text{const.} + \frac{z^2}{2L} - \frac{|z|}{2}.
\end{equation}

Inside the region occupied by the charged sheets, $-w/2 \le z \le
w/2$, we have
\begin{equation}
  \label{eqnSI:phiq-in}
  \phi_q(z) = 4\pi q\bigg(-\frac{zw}{L}+z\bigg).
\end{equation}
Similarly, for $w/2 < z \le L/2$,
\begin{equation}
  \label{eqnSI:phiq-hi}
  \phi_q(z) = 4\pi q\bigg(-\frac{zw}{L} + \frac{w}{2}\bigg),
\end{equation}
while for $-L/2 \le z < -w/2$,
\begin{equation}
  \phi_q(z) = 4\pi q\bigg(-\frac{zw}{L}-\frac{w}{2}\bigg).
\end{equation}

\subsection*{Potential due to a uniformly polarized dielectric}

A uniformly polarized dielectric generates the same electric potential
as a charge distribution comprising two uniformly charged planes,
\begin{equation}
  \rho_{\rm solv}(z) = P\big[\delta_{\rm D}(z-\xi_{\rm hi}) - \delta_{\rm D}(z+\xi_{\rm lo})\big].
\end{equation}
This leads to the following potential,
\begin{equation}
  \phi_{\rm solv}(z) = 4\pi P\bigg[-\frac{z(\xi_{\rm hi} + \xi_{\rm lo})}{L}
    + \frac{\xi^2_{\rm hi} - \xi^2_{\rm lo}}{2L}
    + \frac{1}{2}\big(|z+\xi_{\rm lo}|-|z-\xi_{\rm hi}|\big)\bigg]
\end{equation}

For the region occupied by the dielectric we have ($-\xi_{\rm lo} \le z
\le \xi_{\rm hi}$),
\begin{equation}
  \label{eqnSI:phiP-in}
  \phi_{\rm solv}(z) = 4\pi P\bigg[-\frac{z(\xi_{\rm hi} + \xi_{\rm lo})}{L}
    + z + \frac{\xi^2_{\rm hi} - \xi^2_{\rm lo}}{2L} + \frac{\xi_{\rm lo}-\xi_{\rm hi}}{2}\bigg],
\end{equation}
while for $\xi_{\rm hi} < z \le L/2$
\begin{equation}
  \label{eqnSI:phiP-hi}
  \phi_{\rm solv}(z) = 4\pi P\bigg[-\frac{z(\xi_{\rm hi} + \xi_{\rm lo})}{L}
    + \frac{\xi^2_{\rm hi} - \xi^2_{\rm lo}}{2L}
    + \frac{\xi_{\rm hi}+\xi_{\rm lo}}{2}\bigg],
\end{equation}
and for $-L/2\le z \le -\xi_{\rm lo}$ we have,
\begin{equation}
  \label{eqnSI:phiP-lo}
  \phi_{\rm solv}(z) = 4\pi P\bigg[-\frac{z(\xi_{\rm hi} + \xi_{\rm lo})}{L}
    + \frac{\xi^2_{\rm hi} - \xi^2_{\rm lo}}{2L}
    - \frac{\xi_{\rm hi}+\xi_{\rm lo}}{2}\bigg].
\end{equation}

\subsection*{The total potential}

The total potential is simply the linear superposition of potentials
due to the charged planes and the solvent, $\phi(z) = \phi_q(z) +
\phi_{\rm solv}(z)$. Most important for the derivation is the region
$-\xi_{\rm lo} \le z \le \xi_{\rm hi}$,
\begin{equation}
  \label{eqnSI:phi_inin}
  \phi(z) = 4\pi q\bigg(-\frac{zw}{L}+z\bigg)
  + 4\pi P\bigg[-\frac{z(\xi_{\rm hi} + \xi_{\rm lo})}{L}
    + z + \frac{\xi^2_{\rm hi} - \xi^2_{\rm lo}}{2L} + \frac{\xi_{\rm lo} - \xi_{\rm hi}}{2}\bigg].
\end{equation}
The potential in each of the remaining regions is listed below.
\\ \\
\noindent For $-L/2 \le z < -w/2$:
\begin{equation}
  \label{eqnSI:phi_inoutlo}
  \phi(z) = 4\pi q\bigg(-\frac{zw}{L}-\frac{w}{2}\bigg)
  + 4\pi P\bigg[-\frac{z(\xi_{\rm hi} + \xi_{\rm lo})}{L}
    + \frac{\xi^2_{\rm hi} - \xi^2_{\rm lo}}{2L} - \frac{\xi_{\rm hi} + \xi_{\rm lo}}{2}\bigg].
\end{equation}
For $-w/2 \le z < -\xi_{\rm lo}$:
\begin{equation}
  \phi(z) = 4\pi q\bigg(-\frac{zw}{L}+z\bigg)
  + 4\pi P\bigg[-\frac{z(\xi_{\rm hi} + \xi_{\rm lo})}{L}
    + \frac{\xi^2_{\rm hi} - \xi^2_{\rm lo}}{2L} - \frac{\xi_{\rm hi} + \xi_{\rm lo}}{2}\bigg].
\end{equation}
For $\xi_{\rm hi} < z \le w/2$:
\begin{equation}
  \label{eqnSI:phi_inouthi}
  \phi(z) = 4\pi q\bigg(-\frac{zw}{L}+z\bigg)
  + 4\pi P\bigg[-\frac{z(\xi_{\rm hi} + \xi_{\rm lo})}{L}
    + \frac{\xi^2_{\rm hi} - \xi^2_{\rm lo}}{2L} + \frac{\xi_{\rm hi} + \xi_{\rm lo}}{2}\bigg].
\end{equation}
For $w/2 < z \le L/2$:
\begin{equation}
  \label{eqnSI:phi_outouthi}
  \phi(z) = 4\pi q\bigg(-\frac{zw}{L}+\frac{w}{2}\bigg)
  + 4\pi P\bigg[-\frac{z(\xi_{\rm hi} + \xi_{\rm lo})}{L}
    + \frac{\xi^2_{\rm hi} - \xi^2_{\rm lo}}{2L} + \frac{\xi_{\rm hi} + \xi_{\rm lo}}{2}\bigg].
\end{equation}
Note that $\xi_{\rm hi} + \xi_{\rm lo} = w-\delta$, where $\delta =
\delta_{\rm hi} + \delta_{\rm lo}$.

\subsection*{Linear response}

Equations~\ref{eqnSI:phi_inin}--\ref{eqnSI:phi_outouthi} provide
general expressions for the total electrostatic potential for the
periodic continuum model considered in Fig.~\ref{fig:schematic}. As
$P$ depends upon the electric field, a self-consistent solution is
required. In the case that the dielectric medium is linearly
responding, however, the solution is analytically tractable. Consider
the electric field inside the dielectric. From
Eq.~\ref{eqnSI:phi_inin} we find for $-\xi_{\rm lo} \le z \le \xi_{\rm
  hi}$,
\begin{equation}
  E = -4\pi q\bigg(1 - \frac{w}{L}\bigg) - 4\pi P\bigg(1 - \frac{w-\delta}{L}\bigg).
\end{equation}

Applying the local constitutive relation, $4\pi P = (\epsilon-1)E$, we
find
\begin{equation}
  P = -(\epsilon-1)\bigg[q\bigg(1-\frac{w}{L}\bigg) + P\bigg(1 - \frac{w-\delta}{L}\bigg)\bigg],
\end{equation}
or rearranging,
\begin{equation}
  \label{eqnSI:P-LR}
  P = -\frac{(\epsilon-1)(1-\frac{w}{L})q}{1+(\epsilon-1)(1-\frac{w-\delta}{L})}.
\end{equation}

From Eq.~\ref{eqnSI:phi_inoutlo}, it is clear that the potential at the
charged plate at $z=-w/2$, due to the polarized dielectric is
\begin{equation}
  \label{eqnSI:phisolv_lo}
  \phi_{\rm solv, lo} = 2\pi P\bigg[\frac{w(w-\delta)}{L}
    + \frac{\xi^2_{\rm hi} - \xi^2_{\rm lo}}{L} - (w-\delta)\bigg].
\end{equation}
Similarly, for the charged plate at $z=+w/2$ we have,
\begin{equation}
  \label{eqnSI:phisolv_hi}
  \phi_{\rm solv, hi} = 2\pi P\bigg[-\frac{w(w-\delta)}{L}
    + \frac{\xi^2_{\rm hi} - \xi^2_{\rm lo}}{L} + (w-\delta)\bigg].
\end{equation}
The solvation free energy is $f^{(L)}_{\rm solv} = q(\phi_{\rm solv,
  hi} - \phi_{\rm solv, lo})/2$. Combining
Eqs.~\ref{eqnSI:phisolv_lo}, \ref{eqnSI:phisolv_hi}
and~\ref{eqnSI:P-LR} gives,
\begin{equation}
  f^{(L)}_{\rm solv} =
  -2\pi q^2(w-\delta)\frac{(\epsilon-1)(1-\frac{w}{L})^2}{1+(\epsilon-1)(1-\frac{w-\delta}{L})}.
\end{equation}
In the limit $L\to\infty$ this gives,
\begin{equation}
  \label{eqnSI:fsolvinf}
  f^{(\infty)}_{\rm solv} =
  -2\pi q^2\frac{(w-\delta)(\epsilon-1)}{\epsilon}.
\end{equation}
The finite size correction we must apply is $\Delta f_{\rm DCT}(L) =
f^{(\infty)}_{\rm solv} - f^{(L)}_{\rm solv}$. Thus,
\begin{equation}
  \Delta f_{\rm DCT}(L) = 2\pi
  q^2(w-\delta)(\epsilon-1)\Bigg[\frac{(1-\frac{w}{L})^2}{1+(\epsilon-1)(1-\frac{w-\delta}{L})}
    -\frac{1}{\epsilon}\Bigg].
\end{equation}

\section{Sensitivity of $f_{\rm solv,int}^{(\infty)}$ to $\epsilon_{\rm int}$ and $\ell_{\rm int}$}

\begin{figure}[tb]
  \includegraphics[width=8cm]{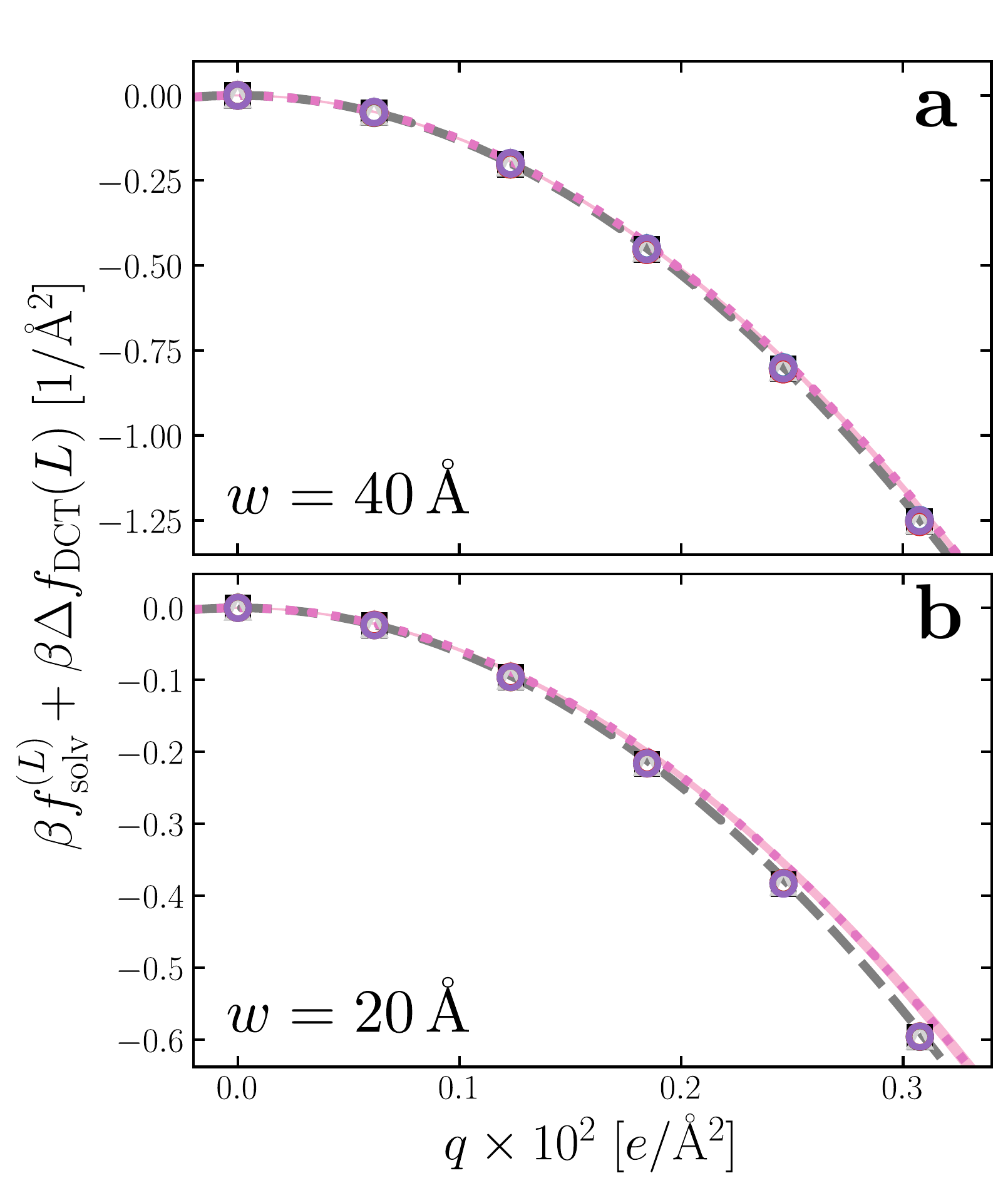}
  \caption{$f_{\rm solv}^{(L)}(q)+\Delta f_{\rm DCT}(L)$ with (a)
    $w=40$\,\AA{} and (b) $w=20$\,\AA. These data are the same as
    shown Figs.~\ref{fig:composite_fesolv}c
    and~\ref{fig:composite_fesolv}d, except $f^{(\infty)}_{\rm
      solv,int}$ (pink dotted lines) is plotted with $\epsilon_{\rm
      int}=10$ and $\ell_{\rm int}=6.0\pm 1.5$\,\AA. While
    discrepancies between $f^{(\infty)}_{\rm solv,int}$ and $f_{\rm
      solv}^{(L)}(q)+\Delta f_{\rm DCT}(L)$ are reduced compared to
    Figs.~\ref{fig:composite_fesolv}c and~\ref{fig:composite_fesolv}d,
    $f^{(\infty)}_{\rm solv}$ given by Eq.~\ref{eqn:fsolvinf} (gray
    dashed lines) still gives a superior description of the simulation
    data.}
  \label{fig:composite_EpsIntTest}
\end{figure}

In Fig.~\ref{fig:composite_EpsIntTest} we plot $f_{\rm
  solv}^{(L)}(q)+\Delta f_{\rm DCT}(L)$ for $w=40$\,\AA{} and
$w=20$\,\AA{} (see Fig.~\ref{fig:composite_fesolv}), but with $f_{\rm
  solv,int}^{(\infty)}$ (Eq.~\ref{eqn:fsolvinf-int}) parameterized
with $\epsilon_{\rm int}=10$ and $\ell_{\rm int}=6.0\pm 1.5$\,\AA. We
argue that $\ell_{\rm int} = \ell_\epsilon \approx 6$\,\AA{} sets a
lower bound on reasonable values of $\ell_{\rm int}$. As discussed in
the main article, increasing $\epsilon_{\rm int}$ and decreasing
$\ell_{\rm int}$, while imposing the constraint $\ell_{\rm
  w}=\ell_{\rm bulk}+2\ell_{\rm int}$ will obviously reduce
discrepancies between $f_{\rm solv,int}^{(\infty)}$ and $f_{\rm
  solv}^{(L)}(q)+\Delta f_{\rm DCT}(L)$, as evidenced by
Fig.~\ref{fig:composite_EpsIntTest}. Nonetheless, it is clear that
$f^{(\infty)}_{\rm solv}$ given by Eq.~\ref{eqn:fsolvinf} still
provides a superior description of the simulation data.

\section{Results with $w=30$\,\AA{} and $w=25$\,\AA{} (WCA centers coincide with the charged planes)}

\begin{figure*}[tb]
  \includegraphics[width=16cm]{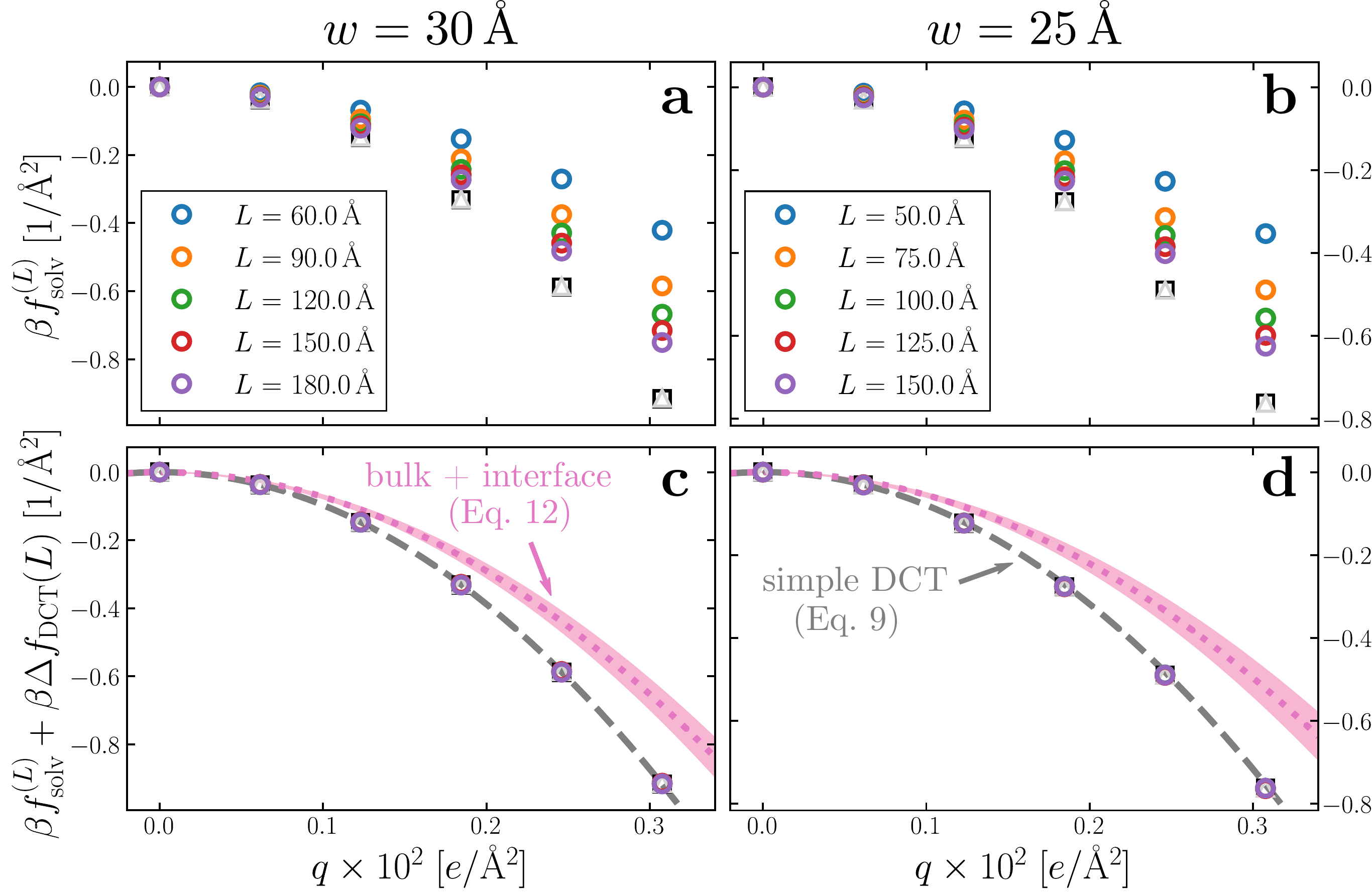}
  \caption{Dependence of solvation free energy $f_{\rm solv}^{(L)}(q)$
    on system size $L$, shown in (a) and (b) for $w=30$\,\AA{} and
    $w=25$\,\AA, respectively.  The values of $L$ for $w=30$\,\AA{}
    are are indicated in the legend of panel (a); those for the
    thinner liquid slab are shown in (b). \tcb{In both cases the WCA
      particles coincide with the charged planes.} Adding \tcb{$\Delta
      f_{\rm DCT}(L)$} given by Eq.~\ref{eqn:fsolvcorr} largely
    removes this sensitivity, as seen in (c) and (d) for $w=30$\,\AA{}
    and $w=25$\,\AA, respectively.  DCT predictions for $f_{\rm
      solv}^{(\infty)}(q)$ (Eq.~\ref{eqn:fsolvinf}) are plotted as
    dashed gray lines. Black squares and gray triangles show results
    obtained with $D=0$\,V/\AA{} for the smallest and largest values
    of $L$, respectively. The pink dotted lines show predictions of
    $f^{(\infty)}_{\rm solv,int}$ from a dielectric continuum model,
    in which an interfacial layer of width $\ell_{\rm int} =
    7.5$\,\AA{} is assigned a permittivity $\epsilon_{\rm int}=2.1$
    much lower than in bulk liquid, computed from
    (Eq.~\ref{eqn:fsolvinf-int}).  The shaded regions bound
    predictions with $6\,{\rm\AA} \le \ell_{\rm int} \le
    9\,{\rm\AA}$.}
  \label{fig:composite_fesolv_suppl}
\end{figure*}

\tcb{In Fig.~\ref{fig:composite_fesolv_suppl} we present results for
  $f_{\rm solv}^{(L)}(q)$ and $f_{\rm solv}^{(L)}(q)+\Delta f_{\rm
    DCT}(L)$ obtained with $w=30$\,\AA{} and $w=25$\,\AA, where in
  both cases, the positions of the WCA particles coincide with the
  charged planes. We draw the same conclusions as from
  Fig.~\ref{fig:composite_fesolv} in the main article.}

\begin{figure}[tb]
  \includegraphics[width=8cm]{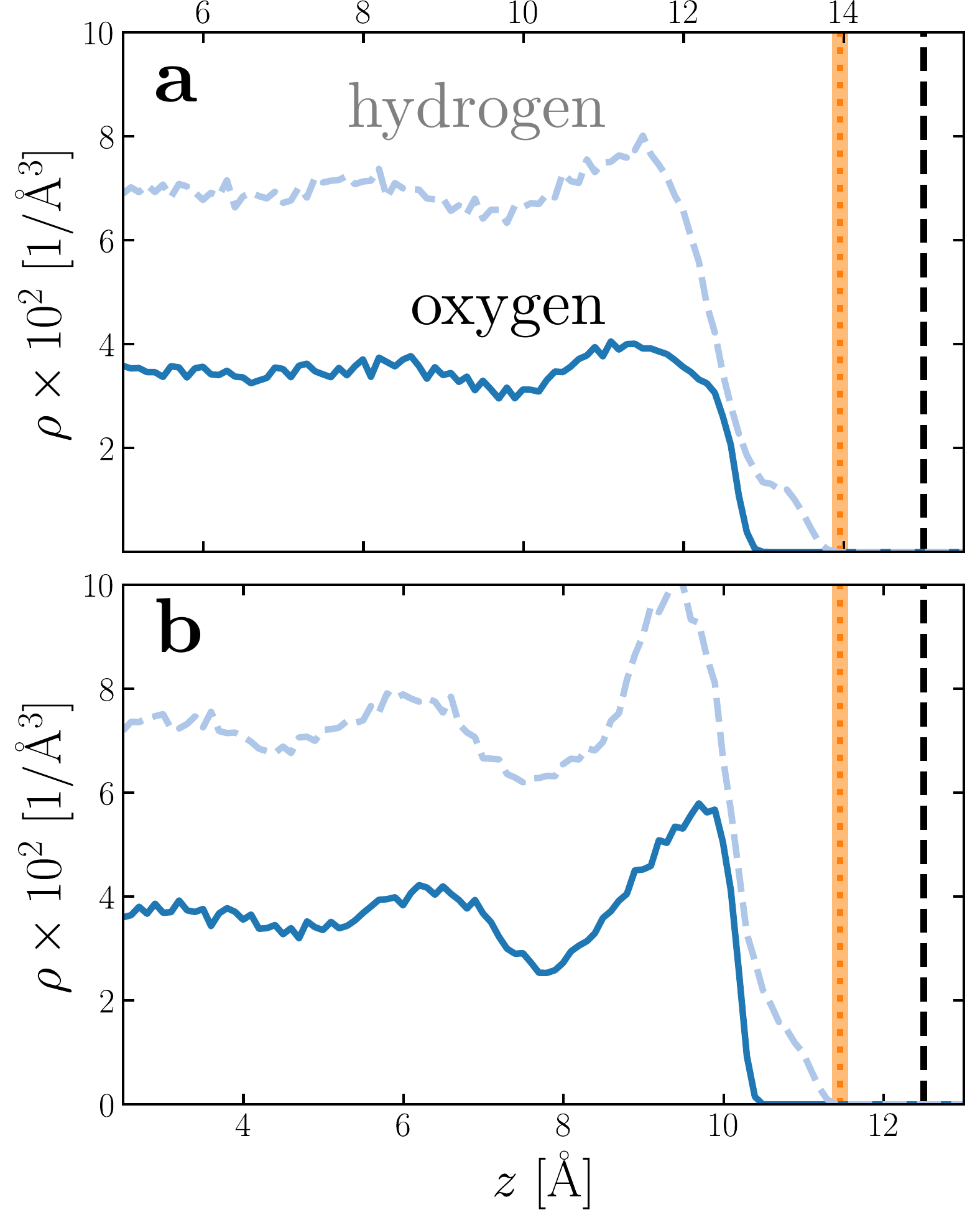}
  \caption{Number density profiles $\rho(z)$ for hydrogen (dashed
    blue) and oxygen (solid blue) atoms of water, with
    $q=0$\,$e$/\AA$^2$ for (a) $w=30$\,\AA{} and (b)
    $w=25$\,\AA. \tcb{In both cases the WCA particles coincide with
      the charged planes.}  The vertical dashed line shows the
    location $z=w/2$ of WCA particles, and the vertical dotted line
    indicates the dielectric boundary at $z=(w-\delta)/2$. (The shaded
    region indicates the same 95\,\% confidence interval as in
    Fig.~\ref{fig:composite_PhiSolvDel}.) In both cases, the
    dielectric boundary aligns closely with the vanishing of hydrogen
    atom density.}
  \label{fig:composite_DensDel_suppl}
\end{figure}

\end{document}